\pdfoutput=1
\documentclass[fleqn,usenatbib]{mnras}
\usepackage{newtxtext,newtxmath}

\usepackage[T1]{fontenc}
\usepackage{ae,aecompl}
\usepackage{times}

\usepackage{graphicx}	
\usepackage{url}
\usepackage{amsmath}	
\usepackage{amssymb}	
\usepackage{setspace}     
\usepackage{flushend}      
\usepackage{etoolbox}      

\makeatletter
\patchcmd\@combinedblfloats{\box\@outputbox}{\unvbox\@outputbox}{}{\errmessage{\noexpand patch failed}}
\makeatother

\newcommand{\eblong}{CRTS J055255.7$-$004426}
\newcommand{\eb}{J0552$-$0044}
\newcommand{\halpha}{H\,$\alpha$}
\newcommand\masyr{mas\,yr$^{-1}$}
\newcommand\kms{\,km\,s$^{-1}$}
\newcommand\Msun{\hbox{$\rm{M}_{\odot}$}}
\newcommand\Lsun{\hbox{$\rm{L}_{\odot}$}}
\newcommand\Rsun{\hbox{$\rm{R}_{\odot}$}}
\newcommand{\vsini}{\ensuremath{v \sin i}}
\newcommand{\gaia}{\emph{Gaia}}
\newcommand{\tess}{\emph{TESS}}
\newcommand{\ellc}{\textsc{ellc}}
\newcommand{\eleanor}{\textsc{eleanor}}

\newcommand{\ebperiod}{0.858968}
\newcommand{\ebvsys}{24.2}
\newcommand{\ebmp}{\ensuremath{0.497\pm0.005}}
\newcommand{\ebms}{\ensuremath{0.205\pm0.002}}
\newcommand{\ebrp}{\ensuremath{0.659\pm0.003}}
\newcommand{\ebrs}{\ensuremath{0.424\pm0.002}}
\newcommand{\ebsb}{\ensuremath{0.520\pm0.001}}
\newcommand{\ebeccnoerr}{0.003}
\newcommand{\ebecc}{\ensuremath{0.003^{+0.001}_{-0.002}}}
\newcommand{\ebarsun}{\ensuremath{3.38\pm0.01}}
\newcommand{\ebaau}{0.016}
\newcommand{\ebloggp}{\ensuremath{4.496\pm0.003}}
\newcommand{\ebloggs}{\ensuremath{4.496\pm0.004}}
\newcommand{\ebteffp}{\ensuremath{3293\pm13}}
\newcommand{\ebteffs}{\ensuremath{3006\pm9}}
\newcommand{\ebloglp}{\ensuremath{-1.337\pm0.013}}
\newcommand{\eblogls}{\ensuremath{-1.879\pm0.013}}
\newcommand{\eblump}{0.046} 
\newcommand{\eblums}{0.013} 
\newcommand{\ebdistbtsettl}{\ensuremath{100\pm5}}
\newcommand{\ebdistphoenix}{\ensuremath{95\pm3}}
\newcommand{\ebvsinicalc}{\ensuremath{38.7\pm0.2}}
\newcommand{\ebvsiniobs}{\ensuremath{37.6\pm0.6}}

\defcitealias{Bell17}{B17}
\defcitealias{Baraffe15}{BHAC15}

\title[Young, low-mass eclipsing binary]{THOR 42: A touchstone $\sim$24 Myr-old eclipsing binary spanning the fully-convective boundary}

\author[S. J. Murphy et al.]{Simon J. Murphy,$^{1,2}$\thanks{E-mail: s.murphy@adfa.edu.au} 
Warrick A. Lawson,$^{1}$ 
Christopher A. Onken,$^{2,3}$ 
David Yong,$^{2,4}$ \newauthor
Gary S. Da Costa,$^{2}$ 
George Zhou,$^{5}$ 
Eric  E. Mamajek,$^{6,7}$  
Cameron P. M. Bell,$^{8}$ \newauthor
Michael S. Bessell$^{2,4}$  
and Adina D. Feinstein$^{9\thanks{NSF Fellow}}$\\\\
$^{1}$School of Science, The University of New South Wales, Canberra, ACT 2600, Australia\\
$^{2}$Research School of Astronomy \& Astrophysics, The Australian National University, Canberra, ACT 2611, Australia\\
$^{3}$Australian Research Council Centre of Excellence for All-Sky Astrophysics (CAASTRO)\\
$^{4}$Australian Research Council Centre of Excellence for All-Sky Astrophysics in 3D (ASTRO 3D)\\
$^{5}$Harvard-Smithsonian Center for Astrophysics, 60 Garden Street, Cambridge, MA, 02138, USA\\
$^{6}$Jet Propulsion Laboratory, California Institute of Technology, 4800 Oak Grove Dr., Pasadena, CA 91109, USA\\
$^{7}$Department of Physics \& Astronomy, University of Rochester, 500 Wilson Blvd., Rochester, NY 14627, USA\\
$^{8}$Leibniz Institute for Astrophysics Potsdam (AIP), An der Sternwarte 16, D-14482 Potsdam, Germany\\
$^{9}$Department of Astronomy \& Astrophysics, University of Chicago, 5640 S. Ellis Ave, Chicago, IL 60637, USA
}

\volume{accepted}
\date{Accepted 2019 November 13. Received 2019 November 13; in original form 2019 October 4}

\pubyear{2019}

\begin{document}
\label{firstpage}
\pagerange{\pageref{firstpage}--\pageref{lastpage}}
\maketitle

\begin{abstract}
\setstretch{1.1}
We present the characterization of \eblong\ (=THOR 42), a young eclipsing binary comprising two pre-main sequence M dwarfs (combined spectral type M3.5). This nearby (103 pc), short-period (0.859 d) system was recently proposed as a member of the $\sim$24~Myr-old 32~Orionis Moving Group. Using ground- and space-based photometry in combination with medium- and high-resolution spectroscopy, we model the light and radial velocity curves to derive precise system parameters. The resulting component masses and radii are \ebmp\ and \ebms\,\Msun, and \ebrp\ and \ebrs\,\Rsun, respectively. With mass and radius uncertainties of $\sim$1 per cent and $\sim$0.5 per cent, respectively, THOR~42 is one of the most precisely characterized pre-main sequence eclipsing binaries known.  Its systemic velocity, parallax, proper motion, colour--magnitude diagram placement and enlarged radii are all consistent with membership in the 32~Ori Group. The system provides a unique opportunity to test pre-main sequence evolutionary models at an age and mass range not well constrained by observation.  From the radius and mass measurements we derive ages of 22--26~Myr using standard (non-magnetic) models, in excellent agreement with the age of the group. However, none of the models can simultaneously reproduce the observed mass, radius, temperature and luminosity of the coeval components. In particular, their H--R diagram ages are 2--4 times younger and we infer masses $\sim$50~per cent smaller than the dynamical values.
\end{abstract}




\section{Introduction}

Understanding the evolution of low-mass stars is an important goal of stellar astrophysics, both observationally and from a theoretical perspective.  Not only because M dwarfs constitute the majority of stars in the Solar neighbourhood \citep[$\sim$80~per cent;][]{Henry04} but, as they have main sequence lifetimes in excess of a Hubble-time, they can also be used to trace the evolution of stellar properties and the structure of the Galaxy \citep[e.g.][]{Reid95,West11}. Furthermore, their prevalence means they are by definition typical planet hosts and prime targets for ongoing transiting planet searches, for instance by the \emph{Transiting Exoplanet Survey Satellite} mission \citep[\tess;][]{Ricker15}.  

However, despite their ubiquity and utility, the properties of low-mass stars (and by extension any planets orbiting them) remain poorly understood, particularly below $\sim$0.35\,\Msun\ where main sequence stars should have fully-convective interiors \citep{Chabrier97}.  With a few exceptions \citep[e.g. angular diameters from interferometry for the nearest M dwarfs;][]{Boyajian12}, many of the fundamental properties of low-mass stars are inferred from comparison to evolutionary models, which tabulate stellar parameters  (most notably radius, temperature and luminosity) as a function of time at a given mass and chemical composition  \citep[e.g.][]{Tognelli11,Bressan12,Baraffe15,Choi16}. For example, the masses and ages of young open cluster and moving group members are almost always inferred from comparison to theoretical tracks and isochrones in the Hertzsprung--Russell (H--R) diagram. These cluster ages set the time-scales for the dissipation of circumstellar discs and the formation of planetary systems \citep[e.g.][]{Mamajek09,Bell13,Murphy18}.  Similarly, our understanding of the stellar initial mass function depends heavily on masses derived from such models \citep{Bastian10}.  As they provide the theoretical backbone to much of contemporary astrophysics, it is therefore imperative that evolutionary models are tested as rigorously as possible against `touchstone' systems with directly-measured properties \citep{Mann15b}. 

Detached, double-lined eclipsing binaries (EBs) are the gold standard for such work. Through detailed photometry and spectroscopy,  the masses and absolute radii (as well as temperatures and luminosities) of both components can be derived to few percent precision with minimal model assumptions \citep{Andersen91,Torres10b}. EBs therefore provide one of the strongest observational tests of theoretical stellar evolution models available \citep{Stassun14,Feiden15b}. Moreover, if an EB is a member of a well-characterized cluster or group, then the models can be calibrated as a function of age. This is especially important at low masses ($\lesssim$0.8\,\Msun), where stellar radii contract rapidly with time as stars descend their Hayashi tracks onto the main sequence. Conversely, the dependence of radius on age means EBs can be used in conjunction with models to age-date a stellar population independently of H--R diagram position \citep[e.g.][]{David19}.  

However, despite recent discoveries, particularly from the \emph{Kepler} $K2$ mission \cite[e.g.][]{Kraus15,Lodieu15,David16a,David16,Gillen17,David19},  the census of young EBs remains small.  There are currently only 16 known low-mass ($<$0.8\,\Msun) systems of age $<$1~Gyr with measured masses and radii for both components, of which only nine are younger than the Pleiades \citep[see compilations by][]{Stassun14,Gillen17,David19}. The majority of these youngest systems are members of Upper Scorpius (age 5--10~Myr) or the Orion Nebula Cluster (ONC; 1--2~Myr).  Moreover, at the lowest masses many systems remain only coarsely characterized, with errors on their component masses or radii of up to $\sim$40 per cent.

\begin{table}
\hfil%
\begin{minipage}{\linewidth}
   \caption{Properties of \eb. Component parameters derived from the light curve and radial velocity modelling are given in \autoref{table:parameters} and photometry is provided in \autoref{table:photometry}.}
   \begin{tabular}{lll} 
\hline
Property & Value & Ref. \\
\hline
Name & \eblong & (1)  \\ 
2MASS & J05525572$-$0044266 & (2)\\
\gaia\ DR2 & 3218460376351485056 & (3) \\
Right Ascension & 05 52 55.732 (ICRS) & (3)  \\
Declination & $-$00 44 27.031 (ICRS) & (3) \\
RA proper motion ($\mu_{\alpha}\cos\delta$) & $11.22\pm0.09$ \masyr & (3) \\
Dec. proper motion ($\mu_{\delta}$) & $-20.57\pm0.08$ \masyr & (3) \\
Parallax & $9.69\pm0.06$ mas & (3) \\
Distance & $102.9\pm0.6$ pc & (4) \\
Spectral type & M3.5 (combined) & (5,6) \\
Radial velocity & $\ebvsys\pm0.4$ \kms\ (systemic) & (5) \\
Age & $24\pm4$ Myr (32 Ori Group) & (5,7)\\
\hline
   \end{tabular}
\emph{References:} (1) \citet{Drake14}; (2)  \citet{Skrutskie06}; (3) \emph\ \gaia\ DR2 \citep{Gaia-Collaboration18}; (4) \citet{Bailer-Jones18}; (5) This work; (6) \cite{Briceno19};  (7) \citet{Bell17}.
\label{table:properties}
\end{minipage}
\hfil%
\end{table}

In this work we present the precise characterization of a young, double-lined system to add to this census. \eblong\ (hereafter \eb, \autoref{table:properties}) was first identified as an EB by \citet{Drake14} in their study of periodic variables in the first data release of the Catalina Surveys (CSDR1). From 192 epochs of pseudo-$V$-band photometry, they determined a period of 0.858956~d and classified \eb\ as an Algol-type detached system. Through comparison to models, \citet{Lee15} derived the mass, fractional radius and age of eclipsing systems identified by the Catalina Surveys. For \eb\ they calculated masses of $M_{1}=0.466\pm0.048$\,\Msun\ and $M_{2}=0.445\pm0.052$\,\Msun, respectively, and a poorly-constrained system age of $8\pm24$~Gyr. The mass and age ranges of their model isochrones meant they were insensitive to lower-mass components at pre-main sequence ages.  

\eb\ was proposed as a possible member of the 24~Myr-old 32~Orionis Moving Group \citep{Mamajek07b}  by \citet{Bell17} (hereafter \citetalias{Bell17}) who noted it was lithium-poor, had a kinematic distance of 92~pc and a UCAC4 proper motion only 2.5\,\masyr\  from that expected of a bona fide group member.  In addition to the CSDR1 light curve, they presented seven radial velocity measurements and fitted Keplerian orbits at the period found by \citet{Drake14}. The resulting orbital solution had a moderate but statistically-insignificant eccentricity ($e=0.1\pm0.11$).  Assuming an edge-on inclination, they derived component masses of  $0.438\pm0.058$\,\Msun\ and $0.164\pm0.019$\,\Msun. Given the large photometric errors and poor coverage of the CSDR1 photometry (\autoref{fig:archival}), they made no attempt to model the light curve and derive radii. With good agreement between the fitted systemic velocity of $20.9\pm2.3$\,\kms\ and the $\sim$20\,\kms\ expected of a 32~Ori Group member, \citetalias{Bell17} considered \eb\ a highly likely member pending further velocity measurements and improved photometry.  

We have re-examined the spectra presented by \citetalias{Bell17} and obtained further velocity measurements and photometry covering the full orbit. In Section\,\ref{sec:observations} we review existing photometric observations of \eb\ and describe our follow-up photometry and spectroscopy. In Section\,\ref{sec:analysis} we jointly model the light curves and radial velocities to derive the properties of the system, including high-precision component masses and radii.  In Section\,\ref{sec:discussion} we discuss our findings in the context of other young EBs, re-assess membership of \eb\ in the 32~Ori Group and compare the system properties to predictions of several  evolutionary model grids. We present our conclusions in Section\,\ref{sec:conclusion}. 

\section{Observations}
\label{sec:observations}

\subsection{Prior photometric surveys}

\eb\ has been observed by several multi-epoch all-sky surveys, which provide a long baseline of photometry at a variety of cadences. \citet{Papageorgiou18} recently used the latest Catalina release (CSDR2) to reanalyse stars identified by \citet{Drake14} as detached EBs, including \eb. Their period agrees with the \citeauthor{Drake14} value to within 1.5 s. $V$-band photometry is also available from the All-Sky Automated Survey for Supernovae \citep[ASAS-SN;][]{Shappee14,Kochanek17}. \citet{Jayasinghe19} homogeneously analysed the ASAS-SN light curves of $\sim$412 000 variables from the VSX catalogue \citep{Watson06}, including \eb. They classified the system as a rotational variable (class \texttt{ROT}, probability 51.2 per cent).  \citet{Karim16} presented 60 $V$-band observations of \eb\ (=CVSO 1975) taken over 12 years as part of their ongoing investigation into young stars in Orion. Although the photometry is of poorer quality than Catalina or ASAS-SN, they found a period of 6.08 d which may be an alias (6.08 d/0.8589 d = 7.08). Finally, \citet{Heinze18} included \eb\ in their catalogue of variable stars from the first data release of the Asteroid Terrestrial-impact Last Alert System \citep[ATLAS;][]{Tonry18b}. ATLAS observed \eb\ in two non-standard bandpasses; cyan (4200--6500\,\AA) and orange (5600--8200\,\AA). From these data, \citeauthor{Heinze18} classified \eb\ as a `dubious' variable with a period of 1.717864\,d (twice the \ebperiod\,d found here). 

\begin{figure}
   \centering
   \includegraphics[width=\linewidth]{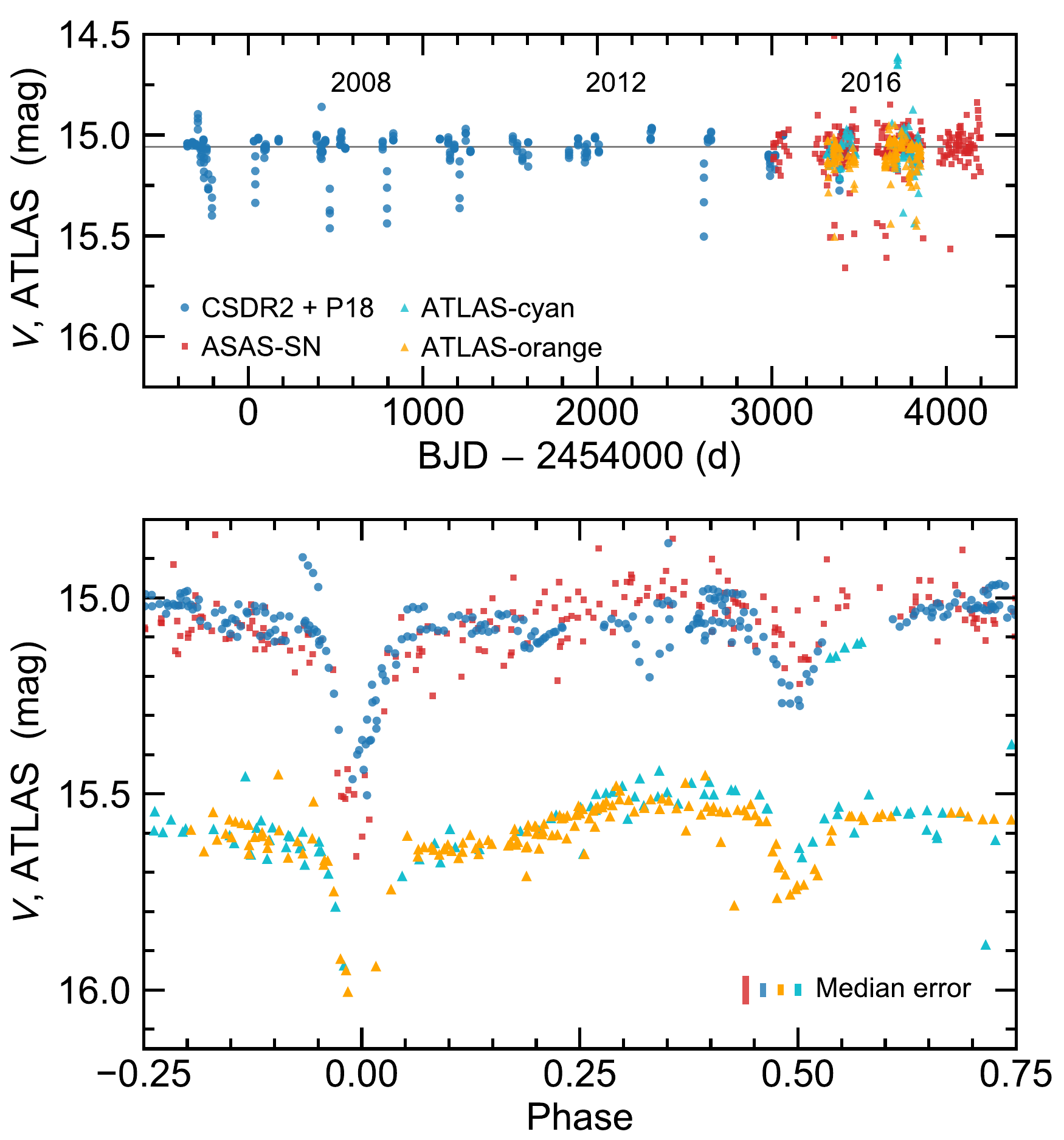}%
   \caption{\eb\ light curves from the Catalina Surveys \citep[CSDR2; blue points, includes additional photometry from][]{Papageorgiou18}, ASAS-SN (red points) and ATLAS DR1 (cyan, orange points) phased to the \ebperiod\,d period found in this work (bottom panel). The Catalina $V_{\rm CSS}$ magnitudes have been shifted to match the median $V$ magnitude from ASAS-SN, while the cyan and orange bandpass ATLAS observations have been shifted by $+0.4$ and $+1.8$~mag, respectively. The combined photometric data span  $\sim$13 yr at a variety of cadences (top panel).}
   \label{fig:archival}
\end{figure}

We plot the CSDR2, ASAS-SN and ATLAS light curves for \eb\ in \autoref{fig:archival}. Both eclipses are visible, with the secondary eclipse around half the depth of the primary. There is also substantial out-of-eclipse variation, with the secondary eclipse occurring at a brighter baseline than the primary. As the secondary eclipse occurs at phase $\phi\approx0.5$, the orbital eccentricity must be close to zero (c.f. \citetalias{Bell17}).  Unfortunately, the large photometric errors and sparse cadence of these all-sky data mean they are ill-suited for fitting light curve models and determining precise radii. We therefore obtained several dedicated photometric data sets (see \autoref{table:photsummary}) to better sample the light curve of \eb, which are described below.

\subsection{Photometry}

\begin{table*}
\hfil%
\begin{minipage}{0.9\linewidth}
   \caption{Summary of \eb\ follow-up photometry.}
   \begin{tabular}{lccccccc} 
\hline
Telescope & Date & Number of & Bandpass & Exposure time & Cadence & Median error \\
 & (\textsc{ut}) & observations & & (s)   & (s) & (mmag) \\
\hline
SkyMapper 1.3-m (SSO) & 2017 Dec 10, 13 & 581 & $i_{\rm SkyMapper}$ & 30 & 50 & 3.8 \\
LCOGT 2-m (Haleakal\=a) & 2018 Jan 14 -- 2019 Jan 7 & 3312 & $i_{\rm LCOGT}$ & 20, 30  & 35, 45 & 1.3\\
\tess\ (Sector 6) & 2018 Dec 15 -- 2019 Jan 6 & 821 & 6000--10\,000\,\AA\ & 1440 & 1800 & 1.8\\
\hline
   \end{tabular}
   \label{table:photsummary}
\end{minipage}
\hfil%
\end{table*}

\begin{table*}
\hfil%
\begin{minipage}{0.96\linewidth}
   \caption{Summary of \eb\ spectroscopic observations.}
   \begin{tabular}{lccccc} 
\hline
Telescope/instrument & Date & Resolution & Wavelength range & Exposure time & S/N$_{\rm H\alpha}^{a}$  \\
& (\textsc{ut}) &  ($\lambda/\Delta\lambda$) & (\AA)  & (s) \\
\hline
ANU 2.3-m/WiFeS (65 epochs) & 2015 Oct 23 -- 2018 Apr 11 & 7000 & 5280--7050 & 1800$^{b}$ & 150\\
 ANU 2.3-m/WiFeS & 2017 Dec 10  & 3000 & 3400--9700 &  2100 & 200\\
 & 2017 Dec 11 &  &  &  1200 & 140\\
Magellan/MIKE (0.7 arcsec slit) & 2017 Nov 27  & 35\,000 (blue) / 27\,000 (red) & 3300--9400 & 300 & 30\\
 & 2017 Nov 28 &   &  & 600 & 40\\
 & 2018 Mar 10 &   &  & 900 & 60\\
 & 2018 Mar 11 &    & & 900  & 60\\
Magellan/MIKE (0.35 arcsec slit)  & 2018 Mar 2  & 83\,000 (blue) / 65\,000 (red) & 3300--9400 & 1800  & 30\\
  & 2018 Mar 4 &   &  & 1200  & 15\\
\hline
   \end{tabular}\\
   $^{a}$Approximate signal-to-noise ratio per pixel measured around H$\alpha$.\\
   $^{b}$Typical exposure time (49/65 exposures). 
      \label{table:spectroscopy}
      \end{minipage}
      \hfil%
\end{table*}

\subsubsection{SkyMapper 1.3-m}

To sample the eclipses at higher cadence and photometric precision we obtained $i$-band observations on the 1.3-m SkyMapper telescope \citep{Wolf18} at Siding Spring Observatory (SSO) on 2017 December 7 (primary eclipse), 10 (secondary) and 13 (primary). Cloud on December 7 prevented the full eclipse from being observed and this noisier partial eclipse was not included in subsequent light curve analysis. The SkyMapper Imager has a field of view of 2.4$^{\circ}$ $\times$ 2.4$^{\circ}$, covered by a 268\,Mpx camera with 0.5 arcsec pixels. On each night we obtained  $\sim$300 $i$-band observations extending 2\,h either side of the predicted eclipse times.  The 30~s exposures and $\sim$20~s readout time gave a median cadence of 50~s. 

The images were processed with a modified version of 
the Science Data Pipeline (SDP) used for Data Release 2 (DR2) of the SkyMapper 
Southern Survey \citep{Onken19}, where the cosmic-ray subtraction 
was deactivated in order to avoid spurious flagging of electronic 
noise peaks. The rest of the 
data reduction proceeded as for DR2: suppression of high-frequency 
interference noise, overscan subtraction, 2D bias subtraction, per-row 
bias structure removed by principal components analysis (PCA), flat-field 
correction, background equalisation between the two amplifiers on each 
CCD, and a PCA-based subtraction of detector fringing. Photometric 
zero-points were based on the ATLAS All-Sky Stellar Reference Catalog 
\citep{Tonry18} after applying Pan-STARRS1-to-SkyMapper bandpass 
transformations \citep[for details see][]{Onken19}.  In a further 
modification from DR2 processing, an individual 
photometric data point consisted of a PSF magnitude determined by a 
2D model created from well-measured stars across each CCD 
using the \textsc{psfex} software package \citep{Bertin11}, where the model was 
allowed to vary linearly with $(x, y)$ position on the CCD.

\begin{figure}
   \centering
   \includegraphics[width=\linewidth]{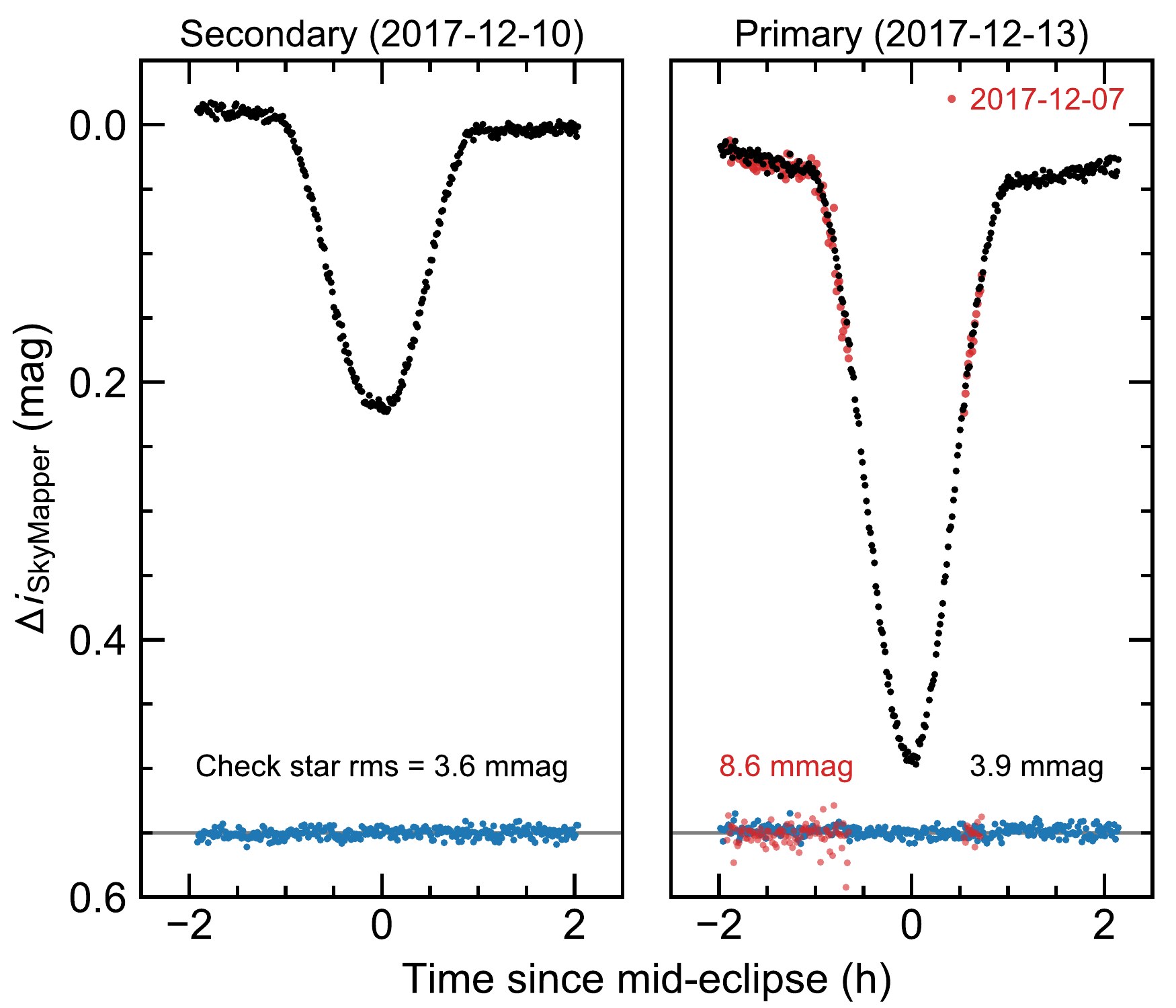}%
   \caption{SkyMapper $i$-band light curves for the secondary eclipse of \textsc{ut} 2017 December 10 (left) and primary eclipse on December 13 (right).  The noisier partial primary eclipse on December 7  (red points) was not used in the analysis.  The check star at the bottom of each panel is the difference between a single comparison star and the mean of the other two.  }
   \label{fig:skymapper}
\end{figure}

We performed differential photometry on these magnitudes using three nearby ($<$2 arcmin) comparison stars of similar brightness to \eb\ to form an unweighted mean comparator and subtracting this from \eb. The resulting eclipse light curves are shown in \autoref{fig:skymapper}.  The typical uncertainty on the differential magnitudes is 3.8\,mmag, which is dominated by a 3.3\,mmag error floor on the individual detections from the SDP. As demonstrated by the pseudo-check star, the rms variation of the comparators is $\lesssim$4~mmag, in agreement with the formal uncertainties.

It is apparent from \autoref{fig:skymapper} that the two eclipses are somewhat asymmetric in that the baseline ingress flux is in both cases higher than egress, with this trend not seen in the check stars across either eclipse.  Although \eb\ is very red ($g-i=2.54$\,mag) compared to most stars in the field (including the comparison stars, see \autoref{fig:tess_tpf}), other stars of similar $(g-i)$ colour in the same exposures show no discernible trend. This seemingly rules out differential atmospheric extinction, which in any case should be small due to the moderate airmass of the observations ($1.15<\sec z < 1.5$).   That a very similar trend appears across a primary and secondary eclipse separated by half an orbit, and is not visible in the other light curves (see below), suggests the cause is observational and not intrinsic to the system (e.g. synchronized spot modulation). In the absence of a physical explanation we mitigate the effect by fitting a quadratic zero-point in time across each night, as discussed in \autoref{sec:fitting}.

\subsubsection{LCOGT 2-m}

\begin{figure} 
   \centering
   \includegraphics[width=\linewidth]{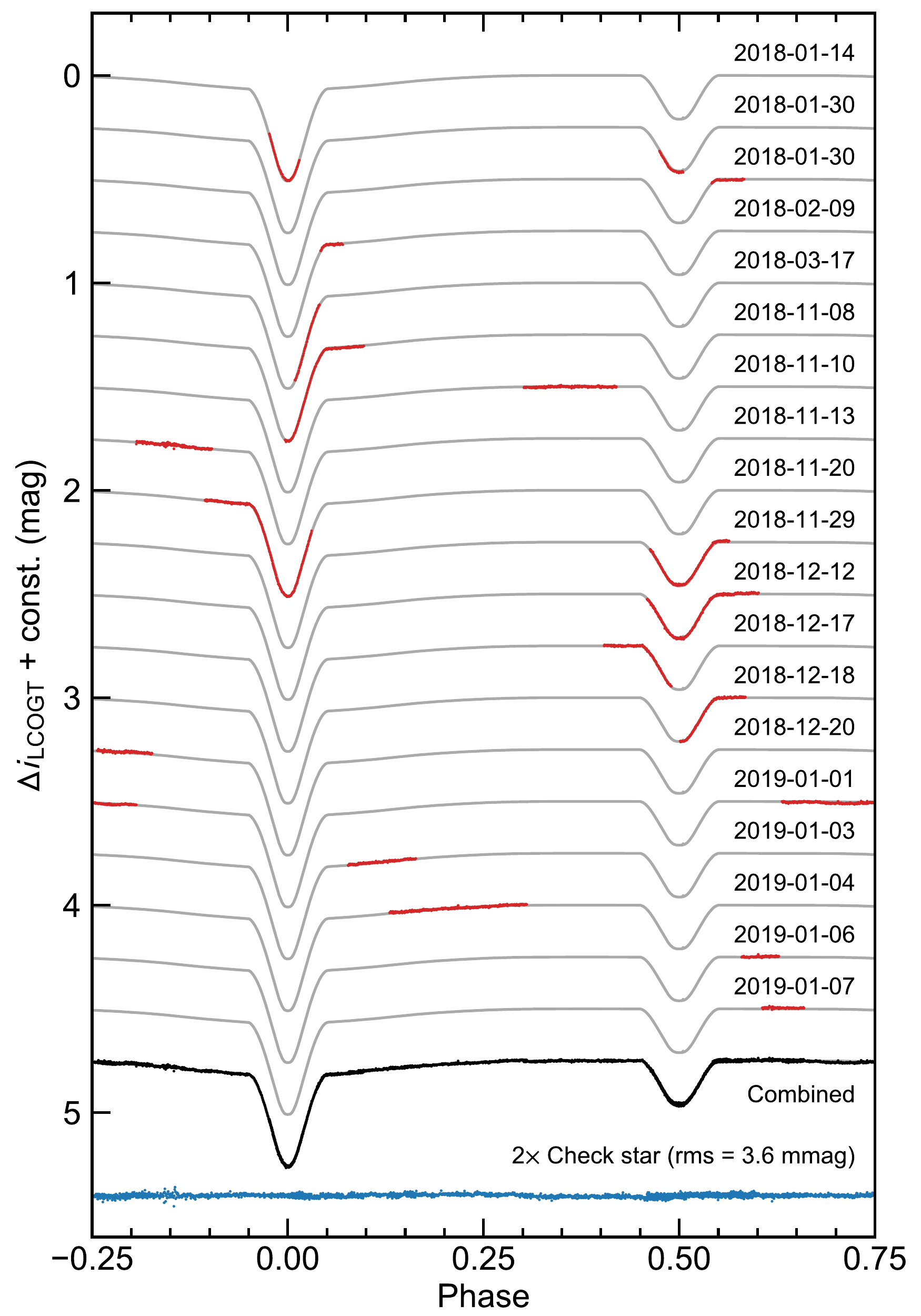}%
   \caption{Phased LCOGT $i$-band observations from 2018/19 (red points). The full light curve is plotted in black; grey lines show the light curve model fitted to the phased data from \autoref{sec:fitting}.  The check star is calculated as in \autoref{fig:skymapper} but has been scaled by a factor of two here for clarity.}
   \label{fig:lcogt}
\end{figure}

We also obtained 40 h of observing time on the Las Cumbres Observatory Global Telescope network \citep[LCOGT;][]{Brown13a},  split between 2\nobreakdash-m telescopes at Haleakal\=a and SSO. Each telescope has a 10.5\,$\times$\,10.5 arcmin field of view and is serviced by a 4096\,$\times$\,4096 pixel imager with 0.3 arcsec pixels at the default 2\,$\times$\,2 binning.  In this work we used only the Haleakal\=a data as it has full phase coverage, better seeing and generally smaller photometric errors than SSO.  We observed for a total of 27 nights from Haleakal\=a, yielding 3809 $i$-band images. Exposure times were 20\,s (30\,s for early 2018 data) with a median cadence of 35 s (45 s).  Dark current, bias and flat-field reduced images were automatically generated by the LCOGT \textsc{banzai} data reduction pipeline \citep{McCully18} at the end of each night using the best-available calibration frames. The pipeline also fits an astrometric solution to every image and extracts object fluxes using an adaptive Kron-like elliptical aperture around each source. Rejecting images with poor transparency, large photometric errors or obvious low-level flares, we retained 3312 observations from 18 nights. We then performed differential photometry on \eb\ as for the SkyMapper images. 

The final LCOGT light curve is plotted in Figs.\,\ref{fig:lcogt} and \ref{fig:best_fit}.  The piecemeal cadence in \autoref{fig:lcogt} is the result of balancing the desire for the longest observing blocks possible against the constraints of the automated LCOGT scheduler and its other high priority targets. Blocks were generally limited to $\lesssim$4\,h duration within a night. The check stars have an rms of 3--4 mmag, significantly more than the typical 1.0--1.5 mmag uncertainties derived by the pipeline.  To ensure we were not underestimating the errors when fitting the light curve model, we fitted a `jitter' term which is added in quadrature to the formal uncertainties before evaluating the model likelihood (see \autoref{sec:fitting} for more information). 

\subsubsection{Transiting Exoplanet Survey Satellite (TESS)}

\begin{figure*}
   \centering
   \includegraphics[width=\textwidth]{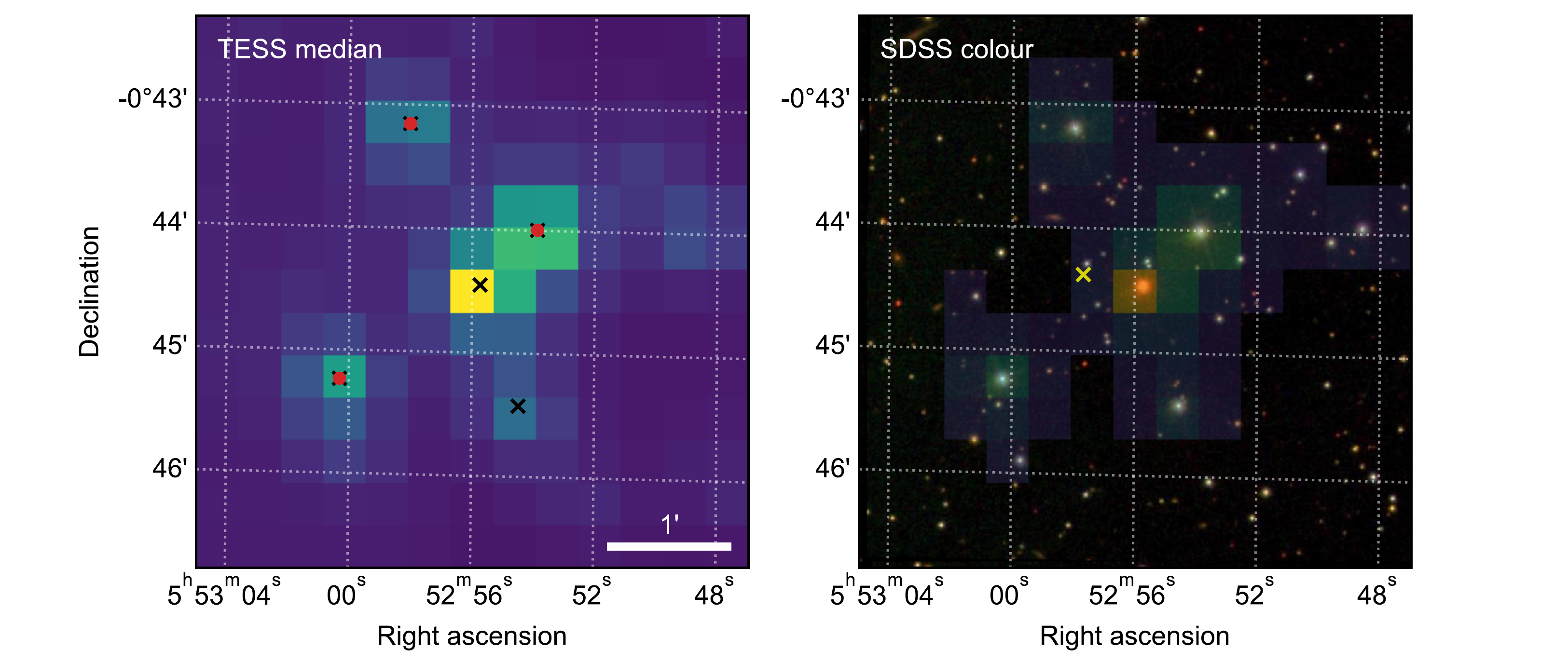}%
   \caption{Left: Median 13\,$\times$\,13  (4.5\,$\times$\,4.5 arcmin) \tess\ Target Pixel File centred on \eb. The four stars with \gaia\ $G$ within 1 mag of \eb\ are shown as crosses; the three used as comparison stars for the SkyMapper and LCOGT photometry are highlighted in red. The fourth star is an unrelated 0.3~d period EB   \citep{Drake14}. Right: Sloan Digital Sky Survey $gri$ colour image of the same field with \tess\ pixels overlaid. The yellow cross denotes the position of the \emph{ROSAT} X-ray source 2RXS J055257.6$-$004422, which we associate with \eb\ (see \autoref{sec:rotation}).}
   \label{fig:tess_tpf}
\end{figure*}

\begin{figure}
   \centering
   \includegraphics[width=\linewidth]{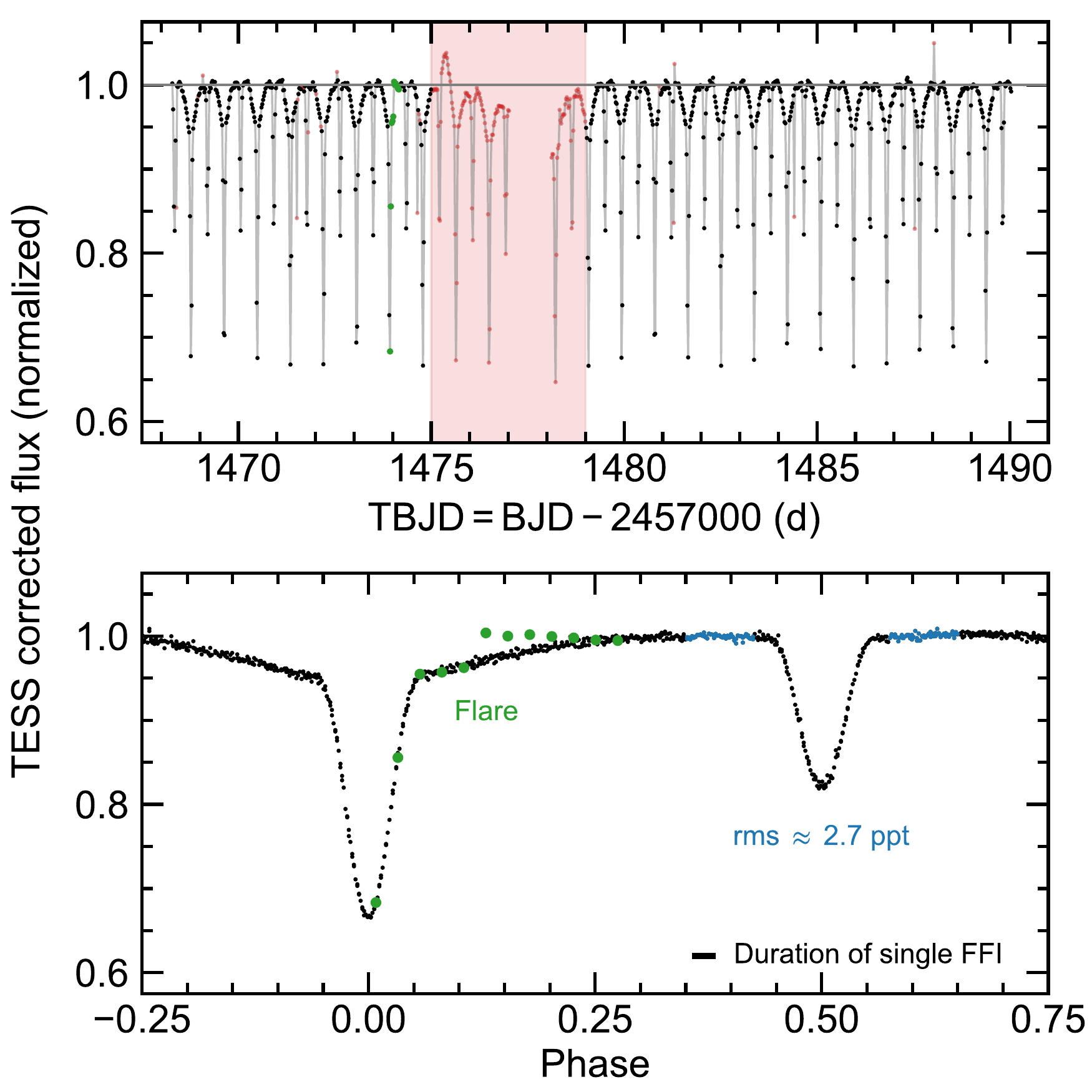}%
   \caption{Normalized \tess\ light curve for \eb. Cadences affected by spacecraft momentum dumps or scattered light (red points) were not included in the normalization or light curve fit. Bottom panel: \eleanor-extracted \tess\ fluxes phased to the \ebperiod\ d period found in this study. The green points show the progression of a flare observed after primary eclipse on 2018 Dec 21, as traced by the 30 min cadence of \tess\ FFIs.}
   \label{fig:tess_lc}
\end{figure}

\eb\ was observed by the \tess\ mission \citep{Ricker15} in Sector 6 of its survey of the southern sky. Observations covered orbits 19 and 20 of the mission between 2018 December 15 and 2019 January 6. As well as 2~min cadence light curves of selected targets, the spacecraft combines 900 consecutive 2~s exposures to create 30~min cadence Full-Frame Images (FFIs) with a scale of 21 arcsec px$^{-1}$ and an effective exposure time of 24~min after onboard cosmic ray mitigation \citep{Vanderspek18}. Sector 6 contains 993 FFIs for each camera/CCD combination. The broad \tess\ bandpass is centred on $\sim$8000\,\AA\ and has a FWHM of approximately 4000\,\AA. 

We used the \eleanor\ software package \citep[v0.2.4;][]{Feinstein19} to download and background-subtract a 13\,$\times$\,13  (4.5\,$\times$\,4.5 arcmin) Target Pixel File centred on \eb\ (\autoref{fig:tess_tpf}) suitable for aperture photometry. While the presence of four bright stars ($\Delta G < 1$~mag) within 1--2 arcmin of \eb\ is useful for differential photometry (e.g.\,SkyMapper, LCOGT), the large \tess\ pixels mean a large aperture will be contaminated by flux from neighbouring stars, diluting the eclipses. We therefore chose to use a custom aperture containing the flux of the brightest central pixel only. The light curve was then corrected in \eleanor\ for systematic trends by regressing against the three most significant co-trending basis vectors (CBVs) for Sector 6, as produced by the \tess\ Science Processing Operations Center (SPOC) pipeline and convolved down from the short cadence data.  

To correct for any remaining systematics and normalize the fluxes, we fitted a low-order polynomial in time to the relatively constant points around secondary eclipse (blue points in \autoref{fig:tess_lc}) for each orbit and divided it into the light curve. Prior to the fit we masked out poor quality cadences ($\texttt{quality} > 0$; provided in \eleanor), images affected by flares (see \autoref{fig:tess_lc}) and any obvious outliers. We also masked out several days at the end of orbit 19 and beginning of orbit 20 (1475\,$<$\,TBJD\footnote{$\mathrm{\tess\ BJD~(TBJD)} = \mathrm{BJD}-2457000$}\,$<$\,1479) where the background changed rapidly as the Earth rose above the spacecraft sunshade.  The resulting light curve has 821 epochs and a median uncertainty of 1.6 parts per thousand (ppt). The standard deviation of fluxes around secondary eclipse is $\sim$2.7~ppt, suggesting the errors provided by the \tess\ pipeline are underestimated for this pixel and a jitter term may be necessary in the light curve fitting.

The SkyMapper, LCOGT and \tess\ observations are summarized in \autoref{table:photsummary}. Time-series photometry  for \eb\ is listed in \autoref{table:lightcurve}. Mid-exposure times are given as Barycentric Julian Dates (BJD) on the Barycentric Dynamical Time (TDB) time-scale.  The normalized \tess\ fluxes have been converted to magnitudes as $-2.5\log(\text{flux})$ and all the light curves shifted so that observations around secondary eclipse have $\Delta m \approx 0.0$.

\begin{table} 
\hfil%
\begin{minipage}{0.57\linewidth}
   \caption{Photometry for \eb\ from SkyMapper, LCOGT and \tess.}
   \begin{tabular}{ccc} 
   \hline
BJD $-$ 2450000 & $\Delta m$ & $\sigma_{m}^{a}$  \\
   (d) & (mag) & (mag) \\
   \hline
\multicolumn{3}{c}{\bf SkyMapper $i$-band}\\
8098.04921 & $-$0.0113 & 0.0038 \\
8098.04980 & $-$0.0128 & 0.0038 \\
8098.05037 & $-$0.0093 & 0.0038 \\
8098.05095 & $-$0.0106 & 0.0038 \\
8098.05152 & $-$0.0160 & 0.0038 \\
\dots & \dots & \dots \\
\multicolumn{3}{c}{\bf LCOGT 2-m $i$-band}\\
8132.89531 & 0.2818 & 0.0011 \\
8132.89584 & 0.2884 & 0.0011 \\
8132.89636 & 0.2972 & 0.0012 \\
8132.89689 & 0.3072 & 0.0012 \\
8132.89742 & 0.3167 & 0.0012 \\
\dots & \dots & \dots  \\
\multicolumn{3}{c}{\bf \tess\ Sector 6}\\
8468.28702 & $-$0.0020 & 0.0018 \\
8468.30784 & 0.0320 & 0.0018 \\
8468.32865 & 0.1696 & 0.0020 \\
8468.34952 & 0.2067 & 0.0021 \\
8468.37034 & 0.0743 & 0.0019 \\
\dots & \dots & \dots \\
\hline
   \end{tabular}\\
   \emph{Note.} This table is published in its entirety in the electronic version of the article. A portion is shown here for guidance regarding its form and content.\\
$^{a}$Does not include additional uncertainties determined in the joint modelling, which should be added in quadrature.
   \label{table:lightcurve}
   \end{minipage}
   \hfil%
\end{table}

\subsection{Spectroscopy}
\label{sec:spectroscopy}

\subsubsection{ANU 2.3-m/WiFeS}

With the aim of deriving radial velocity curves for the system, we obtained 65 epochs of medium-resolution spectroscopy over 25 nights using the Wide Field Spectrograph \citep[WiFeS;][]{Dopita07} on the ANU 2.3-m at SSO. With the $R7000$ grating and $RT480$ dichroic, the spectra cover a wavelength range of 5280--7050\,\AA\ at a resolution of $R\approx7000$. The first seven epochs were presented in \citetalias{Bell17} and the reader is directed to that work and \citet{Murphy15} for details on the observing and data reduction.  The spectra have full phase coverage, with typically several observations per night and an exposure time of 1800~s (see  \autoref{table:spectroscopy}).

\begin{figure} 
   \centering
   \includegraphics[width=\linewidth]{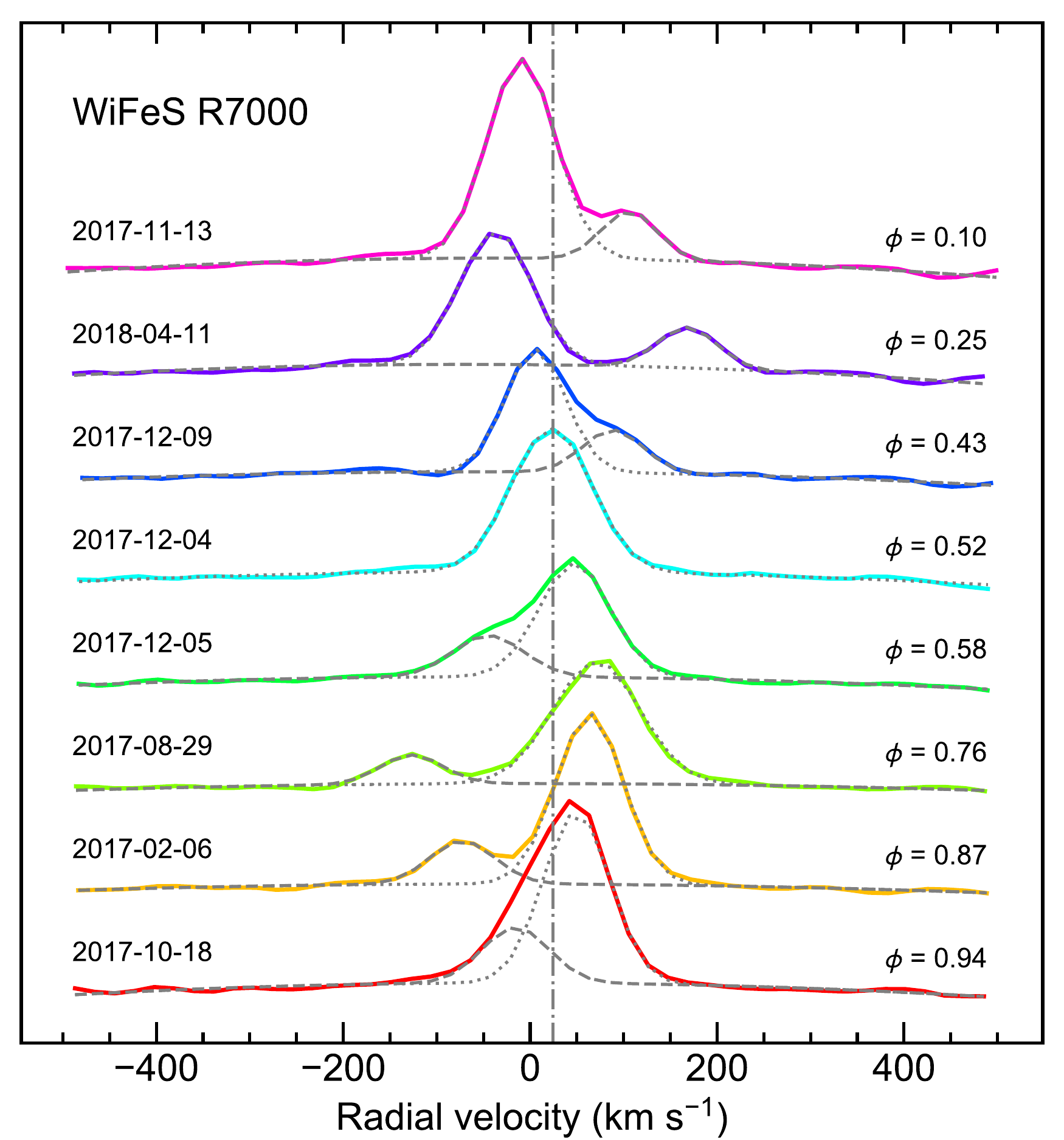}%
   \caption{Selected WiFeS/$R7000$ \halpha\ emission lines with Gaussian fits to the primary (dotted lines) and secondary (dashed lines) stars as a function of phase $\phi$. No secondary was fitted to the $\phi=0.52$ spectrum at secondary eclipse. The vertical line marks the \ebvsys\,\kms\ systemic radial velocity.}
   \label{fig:wifes_halpha}
\end{figure}

\begin{figure*} 
   \centering
   \includegraphics[width=\textwidth]{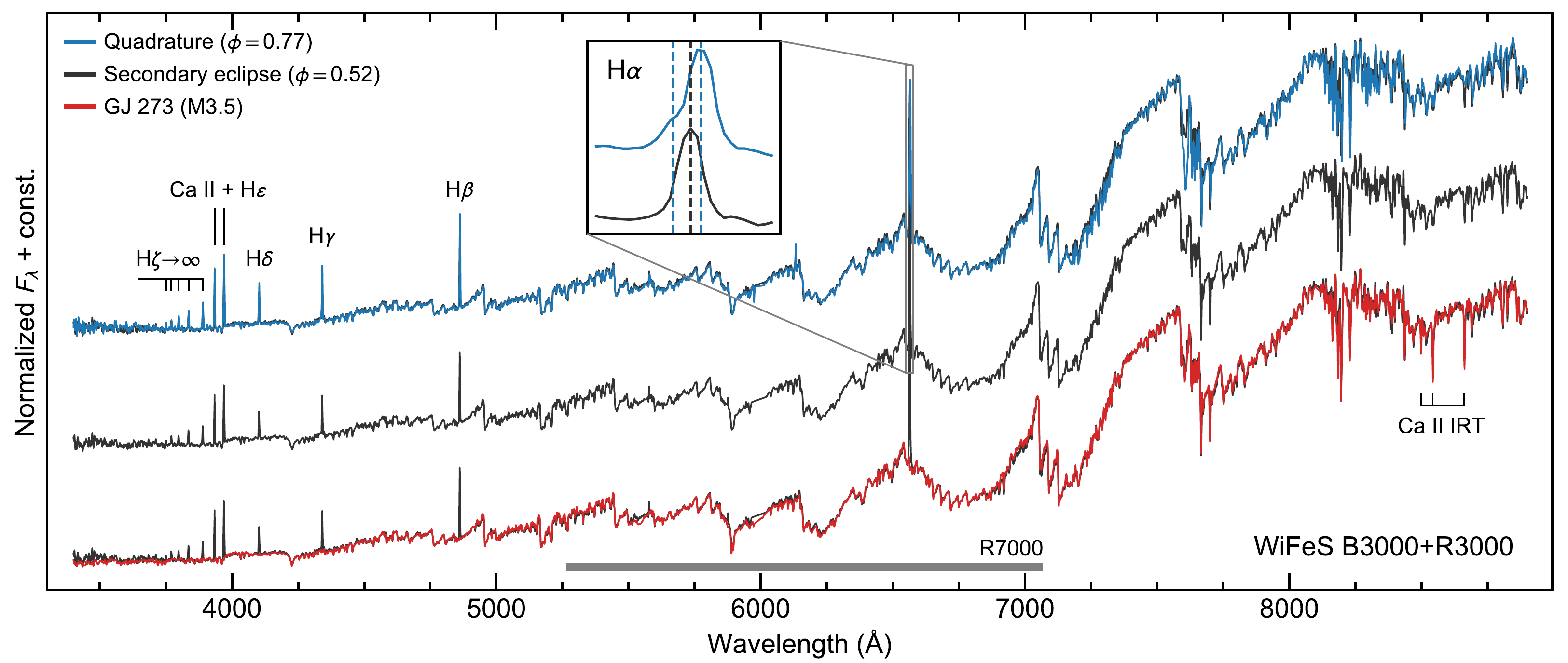}%
   \caption{WiFeS/$B3000+R3000$ spectra of \eb\ at quadrature ($\phi=0.77$) and secondary eclipse ($\phi=0.52$) compared to the M3.5 dwarf GJ~273. All spectra have been normalized over 7400--7550\,\AA. The eclipse spectrum is also plotted under the quadrature and GJ 273 spectra for comparison. The inset shows the change in \halpha\ emission profile between quadrature and eclipse, with dashed lines marking the positions of the components. The grey horizontal bar gives the wavelength coverage of the $R7000$ spectra. Note the much weaker \ion{Ca}{ii} Infrared Triplet (IRT) in \eb\ compared to GJ 273.}
   \label{fig:wifes_3000}
\end{figure*}

As described in \citetalias{Bell17}, we measured radial velocities by cross-correlation against M-type standards from the list of \citet{Nidever02}.  A second set of absorption lines was not visible in either the spectra or cross-correlation functions (CCFs), implying we detected only the primary component.  However, in most cases the \halpha\ emission line was clearly resolved, with a strong component at the primary velocity and a weaker component from the secondary, whose intrinsically stronger emission compensated for the overall flux ratio. We therefore fitted the \halpha\ profile at each epoch with two Gaussians and a quadratic continuum $\pm$500\kms\ around the rest wavelength.  Representative fits are shown in \autoref{fig:wifes_halpha}. We fitted the mean velocity, amplitude and width of each component separately. However, when the separation of the components was minimal ($<$$2\sigma_{\rm tot}$, where $\sigma_{\rm tot}^{2}=\sigma_{\rm pri}^{2} + \sigma_{\rm sec}^{2}$) it was necessary to force both Gaussians to have the same width to avoid over-fitting.

Following this prescription we derived radial velocity differences (RV$_{\rm 2-1}$) for 52/65 epochs. The remaining observations were close to eclipses ($|\textrm{RV}_{\rm 2-1}|\lesssim70$\kms) where only a single component was resolved by WiFeS ($c\Delta\lambda/\lambda\approx45$\kms) or in one case differed by nearly 25\kms\ from the expected value.  These velocities are listed in \autoref{table:rvs}. The primary velocities are the mean and standard deviation against standards observed that run and we have kept the secondary velocities as differences to avoid introducing additional uncertainties. The uncertainty on RV$_{\rm 2-1}$ is the formal error on the mean of each Gaussian added in quadrature. 

We also observed \eb\ using the lower-resolution $B3000$ and $R3000$ gratings and $RT560$ dichroic on \textsc{ut} 2018 December 10 during secondary eclipse ($\phi=0.52$) and the next night near quadrature ($\phi=0.77$). This setup gave coverage of the full optical spectrum (3400--9700\,\AA) at a resolution of $R\approx3000$. We also acquired a spectrum of the M3.5 standard star GJ 273 \citep{Kirkpatrick91} to aid in spectral typing the system. We reduced and combined the blue and red arms in the \textsc{figaro} environment \citep{Shortridge93} using similar techniques to the $R7000$ data and flux calibrated using nightly observations of the spectrophotometric standard L745-46A. The reduced spectra are shown in \autoref{fig:wifes_3000}.  Both epochs of \eb\ are almost identical and are very similar to GJ~273. The only significant differences between the stars are the strong Balmer and \ion{Ca}{ii} H \& K (3969/3934\,\AA) emission compared to GJ 273 and weaker \ion{Ca}{ii} Infrared Triplet (8498/8542/8662\,\AA) absorption, which is presumably filled-in due to activity. The strong similarity to GJ~273 confirms that at optical wavelengths the flux is dominated by the primary component and  hence we assign a (combined) spectral type of M3.5. \citet{Briceno19} recently reported the same spectral type for \eb\ (=CVSO 1975) from their spectroscopic survey of low-mass stars across Orion.

\subsubsection{Magellan/MIKE}
\label{sec:magellan}

\begin{figure} 
   \centering
   \includegraphics[width=\linewidth]{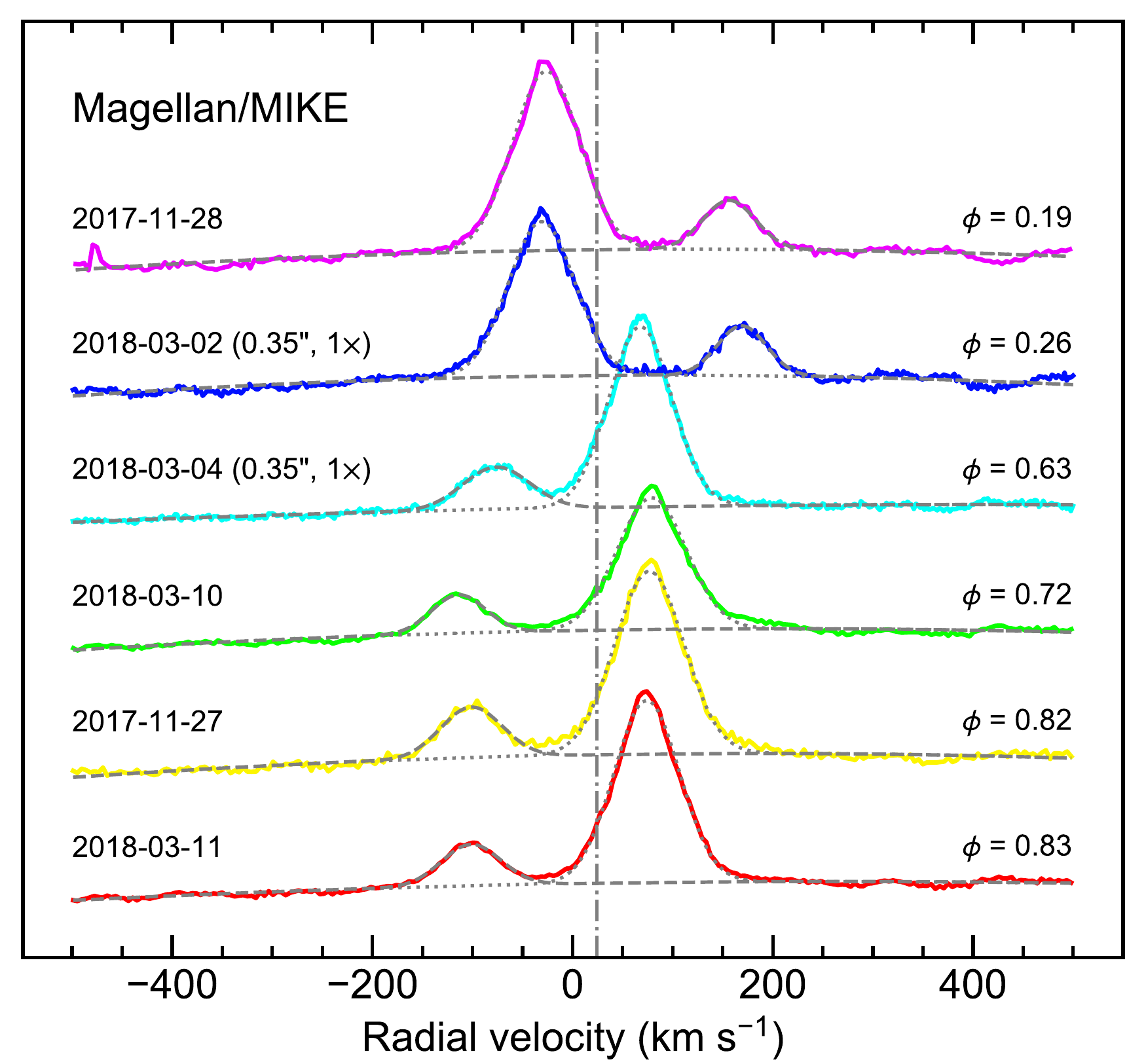}%
   \caption{\halpha\ emission line profiles with Gaussian fits as per \autoref{fig:wifes_halpha}, but for the high-resolution Magellan/MIKE spectra. The two observations taken with the 0.35 arcsec slit and 1$\times$ spectral binning are indicated.}
   \label{fig:mike_halpha}
\end{figure}

Given the short period of \eb, we expect the components to be rotating synchronously (i.e. $P_{\rm rot} = P_{\rm orb} = 0.858$~d). Pre-main sequence models predict synchronous rotation rates of $v\sin i \approx 40$\,\kms\ for the primary component, which would not be resolved by WiFeS. We therefore obtained six observations of \eb\ and GJ 273 using the Magellan Inamori Kyocera Echelle \citep[MIKE;][]{Bernstein03} spectrograph on the 6.5-m Magellan Clay Telescope during 2017 November and 2018 March (see \autoref{table:spectroscopy}).  In most cases we used the $0.7\times5$~arcsec slit ($2\times$ spectral binning, slow readout) which provided a nominal spectral resolution of $R=27\,000$ in the red arm (4800--9400\,\AA) and 35\,000 in the blue (3300--5000\,\AA). However, for two epochs (\textsc{ut} 2018 March 2, March 4) we used the $0.35\times5$~arcsec slit (no spectral binning, fast readout) which gave resolutions of 65\,000 and 83\,000, respectively.  The spectra were wavelength calibrated using contemporaneous Th-Ar arcs. We reduced  the data following standard procedures in the \textsc{carpy} pipeline \citep{Kelson00,Kelson03} to extract and wavelength-calibrate spectra for each of 34 (red) and 36 (blue) echelle orders. No continuum normalization was performed.

We derived radial velocities for the primary component via cross-correlation against each night's spectrum of GJ 273. We did not obtain a spectrum of GJ 273 on \textsc{ut} 2017 November 27 so in this case used the spectrum from November 28. Each spectrograph arm and order was considered separately and the cross-correlation was performed after fitting and subtracting a low-order polynomial continuum from both stars.  Following \citet{White03}, we fitted the peak of the CCF with a Gaussian and quadratic continuum. Given the broad lines seen in \eb, this was a satisfactory model for most orders. The final mean velocity for each arm was calculated after rejecting telluric-dominated orders with velocities $<$5\,\kms\  and performing a 2$\sigma$ iterative clipping to remove outliers.  On the blue side, we did not consider orders $<$4000\,\AA\ due to their lower signal-to-noise ratios and smaller free spectral ranges. Uncertainties for each arm were calculated as the standard deviation across orders. The final velocities at each epoch (see \autoref{table:rvs}) are the weighted mean of the red and blue arms after converting to a barycentric frame and adopting a velocity of $18.22\pm0.1$\,\kms\ for GJ 273 \citep{Nidever02}. Since only a single peak was visible in the CCFs, we calculated the secondary velocities in the same way as the WiFeS spectra by fitting Gaussians to the \halpha\ emission (\autoref{fig:mike_halpha}). \halpha\ is visible on two MIKE orders  so the final secondary velocities ($\rm{RV}_{2-1}$) in \autoref{table:rvs} are the weighted mean of both orders. These velocities agreed to $\lesssim$1.5\,\kms\ for all six observations.

The Gaussian width of each CCF  ($\sigma_{\rm CCF}$) encodes the average width of spectral features in that order, which for \eb\ we assume have been broadened by the fast rotation of the primary compared to GJ 273. Rotation in GJ 273 has not yet been detected, placing an upper limit of $\vsini\leq 2.5$\,\kms\ \citep{Reiners07,Browning10}.  Assuming the only broadening observed in GJ 273 is instrumental, we can therefore directly relate $\sigma_{\rm CCF}$ to \vsini\ \citep{White03}. We determined the calibration between the two quantities empirically for each order by cross-correlating GJ 273 against a spun-up version of itself over the range $5<\vsini<60$\,\kms. The broadened profiles were constructed using the formalism of \cite{Gray08} with a linear limb darkening of $\epsilon=0.6$, and the resulting trend of $\sigma_{\rm CCF}$ versus \vsini\ was well fitted by a cubic polynomial. After excluding orders with velocities outside the calibration range and $\sigma$-clipping as above, we calculated mean red and blue \vsini\ estimates for each observation, which agreed to $<$2\,\kms. Uncertainties in each arm were calculated as the standard deviation across orders.  From these 12 measurements we calculate a weighted mean of $\vsini = \ebvsiniobs$~\kms. The rotation of \eb\ is discussed further in \autoref{sec:rotation}.

\begin{table}
\hfil%
\begin{minipage}{0.94\linewidth}
   \caption{Component radial velocities for \eb\ from ANU \hbox{2.3-m/WiFeS} and Magellan/MIKE spectroscopy.}
   \begin{tabular}{ccccc} 
   \hline
BJD $-$ 2450000 & RV$_{1}$ & $\sigma_{\rm RV_{1}}^{a}$ & $\rm{RV}_{2-1}$  & $\sigma_{\rm{RV}_{2-1}}^{a}$  \\
   (d) & (\kms) & (\kms) & (\kms) & (\kms)\\
   \hline
   \multicolumn{5}{c}{\bf ANU 2.3-m/WiFeS}\\
7319.16766 & 65.5 & 0.7 & $-$150.8 & 1.8 \\
7676.17701 & $-$39.3 & 1.5 & 197.3 & 1.7 \\
7760.99193 & 18.4 & 1.6 &  &  \\
7761.13840 & $-$34.0 & 1.0 &  &  \\
7762.01005 & $-$26.4 & 0.8 & 187.2 & 2.6 \\
\dots & \dots & \dots & \dots & \dots\\
   \multicolumn{5}{c}{\bf Magellan/MIKE (0.7 arcsec slit)}\\
8084.66078 & 76.1 & 0.6 & $-$178.1 & 0.9 \\
8085.83480 & $-$31.1 & 1.1 & 183.2 & 0.8 \\
8188.50503 & 79.9 & 0.7 & $-$192.7 & 0.8 \\
8188.59845 & 74.7 & 0.7 & $-$175.1 & 0.7 \\
   \multicolumn{5}{c}{\bf Magellan/MIKE (0.35 arcsec slit)}\\
8179.52232 & $-$33.0 & 0.7 & 200.2 & 0.5 \\
8181.55586 & 66.4 & 0.6 & $-$144.2 & 0.5 \\
\hline
   \end{tabular}\\
     \emph{Note.} This table is published in its entirety in the electronic version of the article. A portion is shown here for guidance regarding its form and content.\\
$^{a}$$\sigma_{\rm RV_{1}}$ and $\sigma_{\rm{RV}_{2-1}}$ do not include additional uncertainties determined in the joint modelling, which should be added in quadrature.
   \label{table:rvs}
   \hfil%
   \end{minipage}
\end{table}

\section{Analysis}
\label{sec:analysis}

\subsection{Joint light curve and velocity modelling}
\label{sec:fitting}

\begin{table*}
\footnotesize
\hfil%
\begin{minipage}{0.8\linewidth}
   \caption{Fitted and derived parameters from the joint light curve and radial velocity modelling. }
   \renewcommand{\arraystretch}{1.17}
   \begin{tabular}{lccc} 
\hline
Parameter & Symbol & Value & Unit \\
\hline
\multicolumn{4}{c}{\bf Eclipse parameters}\\
Normalized primary radius & $R_{1}/a$ & $0.1951^{+0.0004}_{-0.0005}$ & \\
Normalized secondary radius & $R_{2}/a$ & $0.1254^{+0.0003}_{-0.0004}$ & \\
Orbital inclination & $i$ & $85.84 \pm 0.06$ & $^{\circ}$\\
Eccentricity parameter ($f_{s}$) & $\sqrt{e}\sin\omega$ & $-0.058^{+0.020}_{-0.012}$ & \\
Eccentricity parameter ($f_{c}$) & $\sqrt{e}\cos\omega$ & $0.0020^{+0.0014}_{-0.0010}$ & \\
Orbital period & $P_{\rm orb}$ & $0.85896804 \pm 0.00000007$ & d\\
Reference time of primary eclipse & $t_{0}$ & $8101.13373 \pm 0.00002$ & BJD$-$2450000\\
$i_{\rm LCOGT}$ surface brightness ratio & $J_{\rm LCOGT}$ & $0.520 \pm 0.001$ & \\
$i_{\rm SkyMapper}$ surface brightness ratio & $J_{\rm SkyMapper}$ & $0.541 \pm 0.003$ & \\
\tess\ surface brightness ratio & $J_{\rm \tess}$ & $0.577^{+0.003}_{-0.002}$ & \\
\multicolumn{4}{c}{\bf Light curve parameters}\\
$i_{\rm LCOGT}$ zero-point & $\textrm{ZP}_{\rm LCOGT}$ & $0.00011 \pm 0.00012$ & mag\\
$i_{\rm SkyMapper}$ primary eclipse zero-point (constant) & $\textrm{ZP}_{\textrm{SkyMapper,P0}}$ & $0.017 \pm 0.003$ & mag\\
$i_{\rm SkyMapper}$ primary eclipse zero-point (linear) & $\textrm{ZP}_{\textrm{SkyMapper,P1}}$ & $-0.096 \pm 0.005$ & mag d$^{-1}$\\
$i_{\rm SkyMapper}$ primary eclipse zero-point (quadratic) & $\textrm{ZP}_{\textrm{SkyMapper,P2}}$ & $0.53^{+0.14}_{-0.15}$ & mag d$^{-2}$\\
$i_{\rm SkyMapper}$ secondary eclipse zero-point (constant) & $\textrm{ZP}_{\textrm{SkyMapper,S0}}$ & $-0.00003^{+0.00075}_{-0.00065}$ & mag\\
$i_{\rm SkyMapper}$ secondary eclipse zero-point (linear) & $\textrm{ZP}_{\textrm{SkyMapper,S1}}$ & $-0.094 \pm 0.005$ & mag d$^{-1}$\\
$i_{\rm SkyMapper}$ secondary eclipse zero-point (quadratic) & $\textrm{ZP}_{\textrm{SkyMapper,S2}}$ & $0.69^{+0.18}_{-0.20}$ & mag d$^{-2}$\\
\tess\ zero-point & $\textrm{ZP}_{\rm \tess}$ & $-0.00004^{+0.00022}_{-0.00019}$ & mag\\
$i_{\rm LCOGT}$ jitter & $j_{\rm LCOGT}$ & $3.6 \pm 0.1$ & mmag\\
$i_{\rm SkyMapper}$ jitter & $j_{\rm SkyMapper}$ & $0.5^{+0.5}_{-0.3}$ & mmag\\
\tess\ jitter & $j_{\rm \tess}$ & $3.2 \pm 0.1$ & mmag\\
\tess\ third-light ratio & $\ell_{3,\rm \tess}$ & $0.053 \pm 0.003$ & \\
\multicolumn{4}{c}{\bf Radial velocity parameters}\\
Primary velocity semi-amplitude & $K_1$ & $58.0 \pm 0.4$ & \kms\\
Secondary velocity semi-amplitude & $K_2$ & $140.5 \pm 0.7$ & \kms\\
Systemic velocity (WiFeS) & $v_{\rm sys}$ & $24.2 \pm 0.4$ & \kms\\
Systemic velocity (MIKE) & $v_{\rm sys,MIKE}$ & $23.9^{+0.5}_{-0.6}$ & \kms\\
Secondary velocity offset & $\Delta\textrm{RV}$ & $0.9 \pm 0.5$ & \kms\\
Primary velocity jitter (WiFeS) & $j_{\rm 1}$ & $3.2 \pm 0.3$ & \kms\\
Primary velocity jitter (MIKE) & $j_{\rm 1,MIKE}$ & $0.8^{+1.0}_{-0.5}$ & \kms\\
Secondary velocity jitter (WiFeS) & $j_{\rm 2}$ & $3.5 \pm 0.5$ & \kms\\
Secondary velocity jitter (MIKE) & $j_{\rm 2,MIKE}$ & $1.9^{+1.0}_{-0.7}$ & \kms\\
\multicolumn{4}{c}{\bf Spot parameters}\\
Spot central longitude & $l_{\rm spot}$ & $0.1 \pm 0.1$ & $^{\circ}$\\
Spot central latitude & $b_{\rm spot}$ & $-26^{+2}_{-3}$ &$^{\circ}$\\
Spot radius & $r_{\rm spot}$ & $39 \pm 1$ & $^{\circ}$\\
$i_{\rm LCOGT}$  spot brightness ratio & $B_{\textrm{spot,LCOGT}}$ & $0.81^{+0.01}_{-0.02}$ & \\
$i_{\rm SkyMapper}$ spot brightness ratio & $B_{\textrm{spot,SkyMapper}}$ & $0.82^{+0.01}_{-0.02}$ & \\
\tess\ spot brightness ratio & $B_{\textrm{spot},\tess}$ & $0.82 \pm 0.01$ & \\
\multicolumn{4}{c}{\bf Limb darkening parameters}\\
$i_{\rm LCOGT}$ triangular sampling parameter 1 & $q_{1,{\rm LCOGT}}$ & $0.99^{+0.01}_{-0.02}$ & \\
$i_{\rm LCOGT}$ triangular sampling parameter 2 & $q_{2,{\rm LCOGT}}$ & $0.12 \pm 0.03$ & \\
$i_{\rm SkyMapper}$ triangular sampling parameter 1 & $q_{1,{\rm SkyMapper}}$ & $0.96^{+0.03}_{-0.07}$ & \\
$i_{\rm SkyMapper}$ triangular sampling parameter 2 & $q_{2,{\rm SkyMapper}}$ & $0.20^{+0.05}_{-0.04}$ & \\
\tess\ triangular sampling parameter 1 & $q_{1,{\rm \tess}}$ & $0.74^{+0.10}_{-0.11}$ & \\
\tess\ triangular sampling parameter 2 & $q_{2,{\rm \tess}}$ & $0.05^{+0.08}_{-0.03}$ & \\
\multicolumn{4}{c}{\bf Derived parameters}\\
Primary radius & $R_{1}$ & $0.659^{+0.002}_{-0.003}$ & \Rsun\\
Secondary radius & $R_{2}$ & $0.424 \pm 0.002$ & \Rsun\\
Primary mass & $M_{1}$ & $0.497 \pm 0.005$ & \Msun\\
Secondary mass & $M_{2}$ & $0.205 \pm 0.002$ & \Msun\\
Mass ratio ($M_{2}/M_{1}$) & $q$ & $0.413^{+0.005}_{-0.004}$ & \\
Primary surface gravity & $\log g_{1}$ & $4.496 \pm 0.003$ & cgs\\
Secondary surface gravity & $\log g_{2}$ & $4.496 \pm 0.004$ & cgs\\
Orbital semi-major axis & $a$ & $3.38 \pm 0.01$ & \Rsun\\
Orbital eccentricity & $e$ & $0.003^{+0.001}_{-0.002}$ & \\
Longitude of periastron & $\omega$ & $272^{+2}_{-1}$ & $^{\circ}$\\
\hline
\end{tabular}\\
\label{table:parameters}
\end{minipage}
\hfil%
\end{table*}

Having collected light curves and radial velocities, we modelled the physical parameters of \eb\ and their uncertainties. To accomplish this we used the Python binary light curve package \ellc\ \citep[v1.8.4;][]{Maxted16}\footnote{\url{https://github.com/pmaxted/ellc}}, which represents the components of the binary as triaxial ellipsoids and calculates the observed flux using a combination of Gauss-Legendre integration and exact analytical expressions for the areas of overlapping ellipses which are the projection of these ellipsoids on the sky.   As described by \cite{Maxted16}, the flux is calculated from the visible area of the ellipses, which can be calculated exactly, weighted by the average intensity over the visible area, which is estimated by numerical integration over a Cartesian grid defined by the major and minor axes of each ellipse. For efficiency we considered spherical stars with no gravity darkening and adopted the `\texttt{very\_sparse}' grid which limits the numerical integration to $n=4$ points along each axis. 

Our model comprises 43 parameters to describe the SkyMapper, LCOGT, \tess, WiFeS and Magellan data, which are summarised in \autoref{table:parameters}.   As well as standard parameters describing the stellar radii, disc-averaged surface brightness ratios, reference time, orbital period, eccentricity and inclination, we also included additional parameters to better describe the light curves, radial velocities, star spots and limb darkening, which are detailed below. Note that the orbital semi-major axis, $a$, is parametrized in \ellc\ as:
\begin{align}
a &= a_1\,(1+1/q) \quad\textrm{for mass ratio } q = M_2/M_1\textrm{ and} \\
a_1\sin i &= 0.0197657\,K_1\,P_{\rm orb} \sqrt{1 - e^2}
\end{align}
where ($a, a_1$) are in solar radii, $K_1$ is in \kms, $P_{\rm orb}$ in days, and the numerical factor uses the nominal solar constants given in IAU 2015 Resolution B3 \citep{Mamajek15b}.

\subsubsection{Light curve modelling}
\label{sec:lc_modelling}

We modelled each photometric data set (SkyMapper, LCOGT, \tess) with a zero-point offset and jitter term as free parameters to account for minor normalization differences and underestimated uncertainties. In addition to the constant zero-point term, each SkyMapper eclipse included a quadratic in time to correct the observations for the small trends seen in \autoref{fig:skymapper}.  Because of the large \tess\ pixels and the possibility the extracted flux was contaminated by neighbouring bright stars, we also included a third light component ($\ell_3$) in the \tess\  light curve model \citep[see][]{Maxted18}\footnote{The definition of the third light parameter in the latest versions of \ellc\ is different to the one described in \citet{Maxted16}. See \citet{Maxted18} for more information.}. For efficiency, we evaluated the full light curve model at every fifth point in the LCOGT time-series, making use of \ellc's ability to linearly interpolate the model at intermediate times. No interpolation was performed between nightly light curve segments. Conversely, we integrated each 1440 s \tess\ observation over 10 subsamples in \ellc\ to account for the longer exposure time.

\subsubsection{Radial velocities}

We modelled the radial velocity curves in \ellc\ simultaneously with the light curves as centre-of-mass velocities assuming Keplerian orbits (i.e. ignoring the Rossiter--McLaughlin effect). Secondary velocities were calculated as the RV$_{2-1}$ difference to match the observed \halpha\ offsets in \autoref{table:rvs}. Note that this means in effect that the systemic velocities are calculated solely from the primary component. The model also included jitter terms for each of the four data sets and we integrated each observation over 150 s subsamples to account for the non-negligible exposure time (typically 12 samples for an 1800~s WiFeS exposure). To minimize any systematic differences between the WiFeS and MIKE data sets we fitted each with a separate systemic velocity. Furthermore, to mitigate any biases in the primary velocities (which were derived from cross-correlation of photospheric absorption lines) versus the secondary velocities (which originated from chromospheric \halpha\ emission), we also included an additional offset ($\Delta$RV) on the secondary velocities before they were compared to the observations. 

\subsubsection{Spot modelling}

\eb\ exhibits out-of-eclipse variation, characterized by a broad depression around primary eclipse returning to an approximately flat baseline around secondary eclipse. This is somewhat similar to the reflection/heating effect seen in binaries with hot and cool components, whereby flux from the brighter star (typically an OB star or white dwarf) impinges on the visible disc of the cooler companion (typically an M dwarf). As both components of \eb\ are M dwarfs and should differ in temperature by only a few hundred~K, we do not expect reflection effects to be significant. Nevertheless, we experimented with various combinations of reflection parameters in \ellc\ and could satisfactorily replicate the observed light curves. However, these attempts generally required unphysical radii and significant ellipsoidal and/or gravity darkening effects to match the observations, and yielded poor fits to the radial velocities. For instance, given that reflection effects typically produce concave-up light curves around secondary eclipse, with no gravity darkening the model required $R_2 > R_1$ to generate the large ellipsoidal variations at $\phi=0.25$ and 0.75 necessary to flatten the light curve and replicate the observations. 

In light of these deficiencies, we chose to model the out-of-eclipse variation as resulting from the passage of a single dark spot on the surface of the primary (equivalent to a bright spot on the secondary). Spots can naturally explain the light curve modulation and are endemic on low-mass stars. The spot is parametrized in \ellc\ as a circle with central longitude, latitude, angular radius and brightness ratio ($B_{\rm spot} < 1$ for a dark spot).  We assigned each bandpass its own brightness factor in the model. Admittedly, such a simple model is almost certainly a crude approximation to the complex spot patterns present on both stars and their evolution with time. However, toy models show that the net effect is well-modelled by a single, unchanging spot centred near longitude zero (i.e. in line with the secondary as it orbits, see \autoref{fig:configuration}).  Both stars are magnetically active and should be synchronously rotating (see \autoref{sec:rotation}), so the presence of the companion is likely responsible for the preferential longitude and longevity of the spot pattern.

\subsubsection{Limb darkening}

Rather than constraining the limb darkening coefficients with firm priors or fixing them at values appropriate for the components' temperatures and surface gravities \citep[e.g.][]{Claret11,Parviainen15}, we allowed them to vary as free parameters in the fit.  To ensure the quadratic coefficients $(u_1,u_2)$ remained physically bounded, we transformed them to the $q_1 = (u_1 + u_2)^2$ and $q_2 = 0.5u_1(u_1+u_2)^{-1}$ triangular sampling parametrization proposed by \citet{Kipping13}, with uniform priors on $(q_1,q_2)$ over the interval $[0, 1]$. At each iteration, $(q_1,q_2)$ were transformed back to $(u_1,u_2)$ for use in \ellc.  In this way we could derive posterior probabilities on all parameters which fully accounted for our ignorance of the limb darkening profile, yet never explored unphysical solutions. We additionally required that both components share the same (bandpass-dependent) coefficients. As the stars should have similar surface gravities and  temperatures, this is a reasonable approach which halves the number of coefficients required.
  
\subsubsection{Parameter estimation}

We explored the model parameter space using the Markov Chain Monte Carlo (MCMC) ensemble sampler \textsc{emcee} \citep[v2.2.1;][]{Foreman-Mackey13,Goodman10} to sample the posterior probability distribution of the model and find parameter sets which best describe the observations.   At each iteration we calculated the log-likelihood of model parameters $\theta$ given the observed data sets $\mathbf{x} =  \cup_i x_i$ as follows:
\begin{equation}
\ln\mathcal{L}(\theta | \mathbf{x} ) = -\frac{1}{2}\sum_{i} \chi^2_{i}\quad\textrm{for}\quad i=\textrm{LCOGT, SkyMapper, \dots}
\end{equation}
where the $i$th data set comprises $n$ observations:
\begin{equation}
\chi^2_{i} = \sum_{n} \ln (2\pi\sigma_n^2) + [\rm{obs}_n - \rm{model}(\theta)_n]^2/\sigma_n^{2}
\end{equation}
and the uncertainties include a jitter term, $j_i$, added in quadrature to the observed uncertainties:
\begin{equation}
\sigma_n^{2} = \sigma^{2}_{\textrm{obs}, n} + j_i^{2} 
\end{equation}

We adopted uniform priors on all parameters, imposing physical limits where appropriate. We then sampled the parameter space 100\,000 times in \textsc{emcee} using 128 walkers.  Each walker was initialized from a Gaussian parameter ball around a hand-tuned solution which provided a reasonable fit to the observations.  After confirming convergence by inspecting the parameter  and posterior probability traces, we conservatively discarded the first 90\,000 steps in each MCMC chain as burn-in and report parameter values from the remaining 10\,000 steps. The values in \autoref{table:parameters} are the median and $\pm1\sigma$ Gaussian-equivalent uncertainties formed from the 16th and 84th percentiles of the parameter distributions.

\begin{figure}
   \centering
   \includegraphics[width=\linewidth]{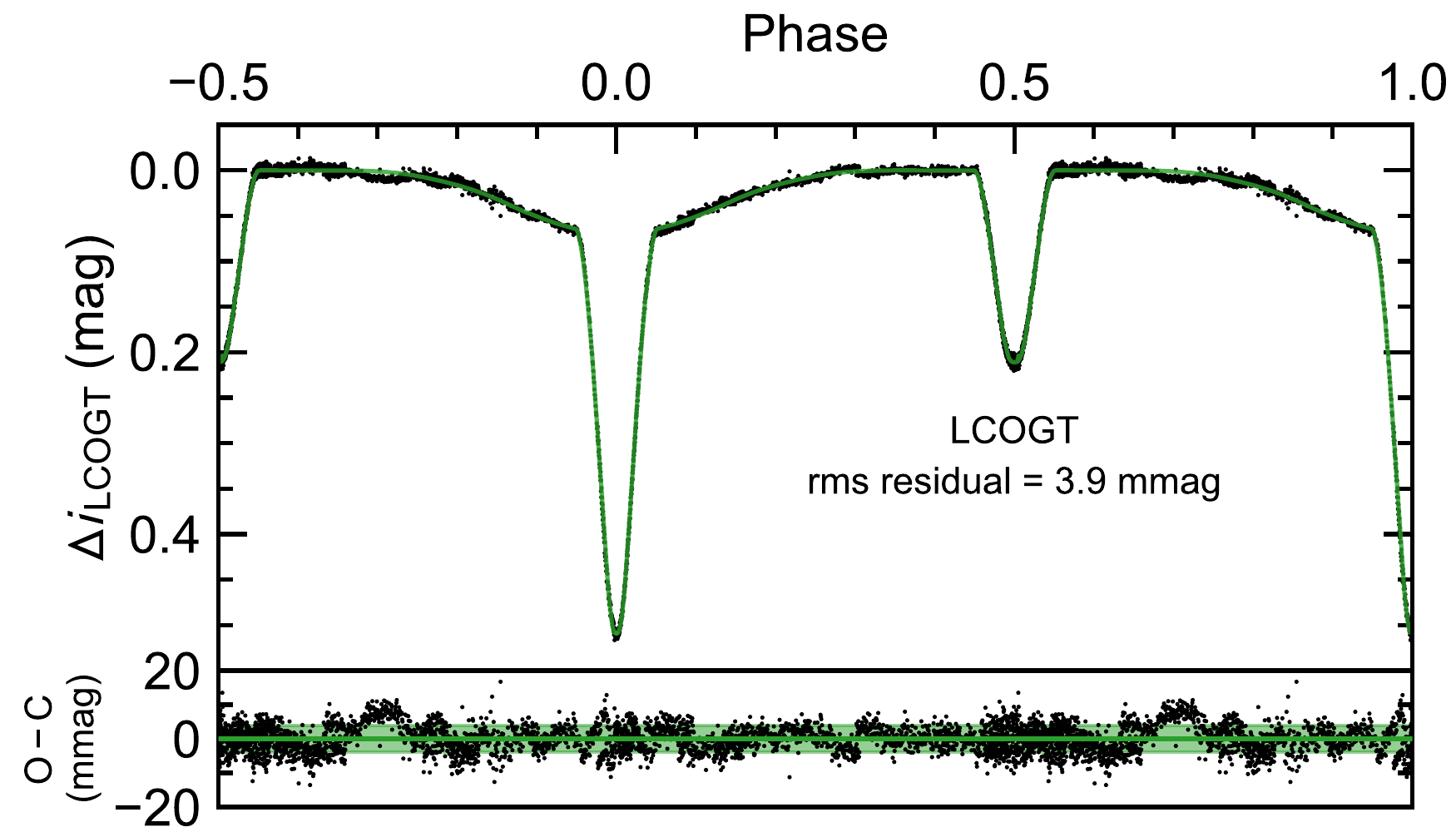}
   \includegraphics[width=\linewidth]{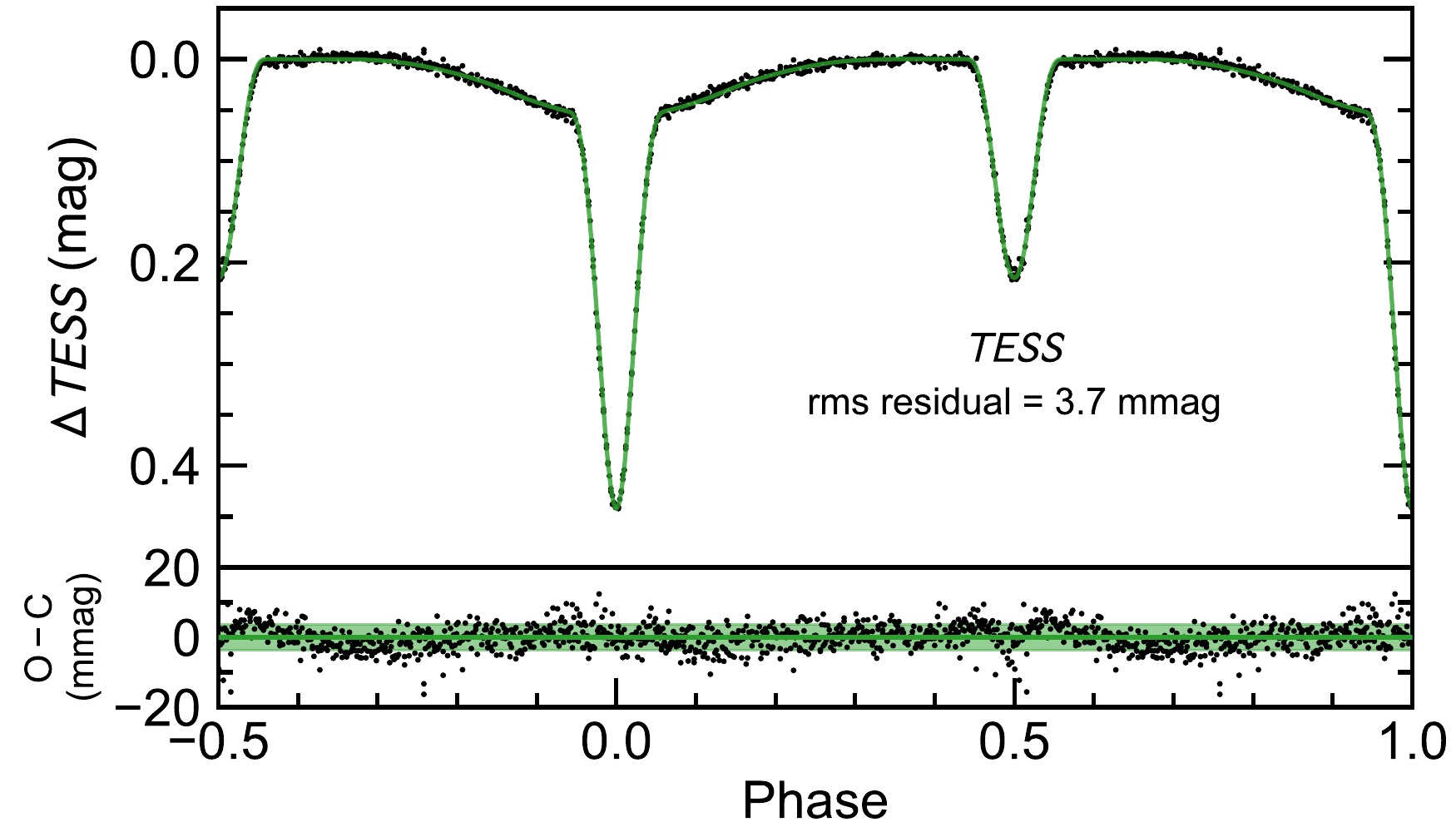}%
   \caption{Phase-folded LCOGT and \tess\ light curves for \eb\ with the best-fitting model shown in green. Fit residuals are shown in the bottom panel of each plot; the green shaded bars show the rms residual.}
   \label{fig:best_fit}
\end{figure}

\begin{figure}
   \centering
   \includegraphics[width=\linewidth]{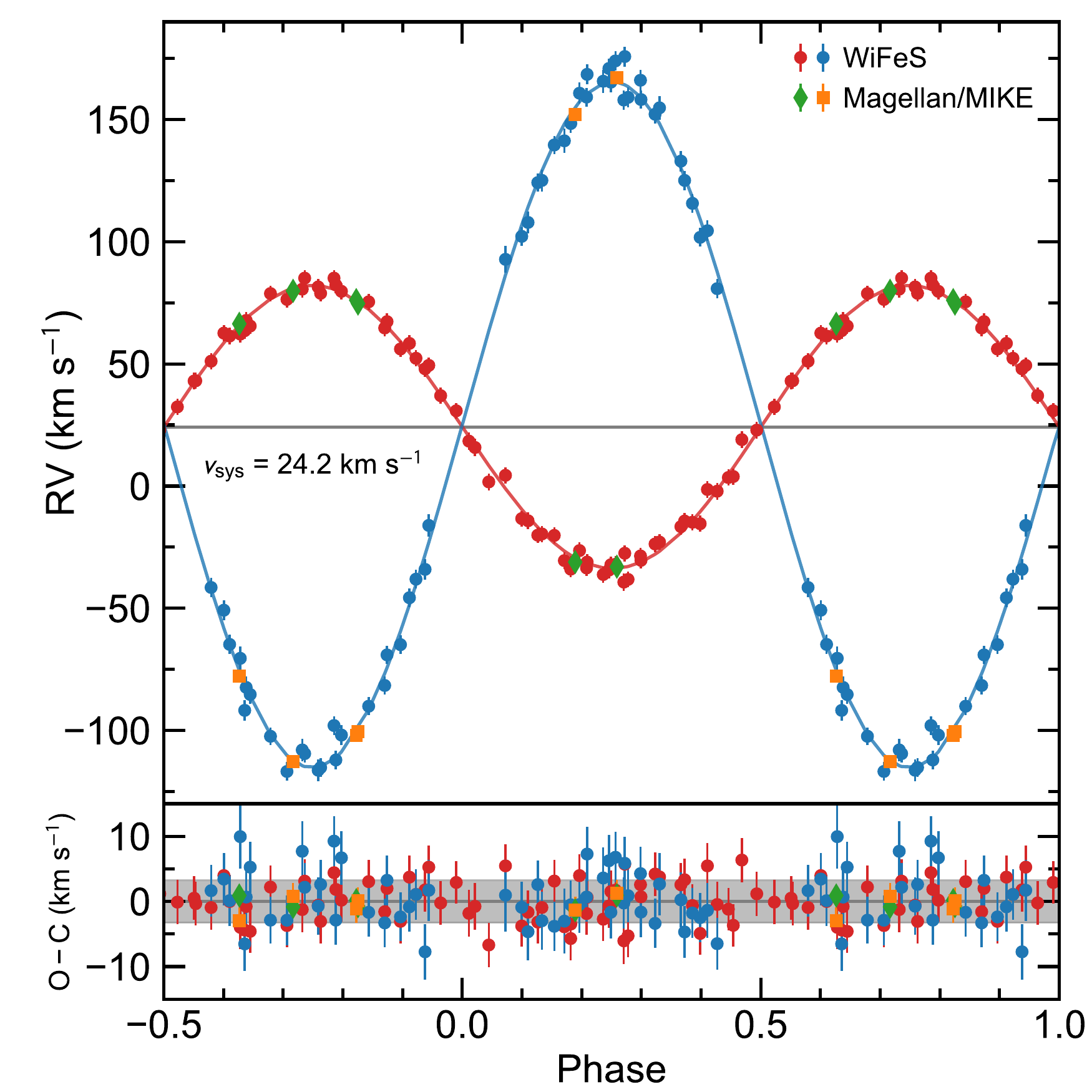}%
   \caption{Radial velocity measurements for \eb\ from WiFeS (red, blue points) and Magellan/MIKE (green, orange points), with the best-fitting orbital solution given by solid lines.  Error bars include the jitter terms added in the fitting. The primary and secondary rms residuals are 3.3 and 4.1\,\kms\ for WiFeS, and 0.9 and 1.5\,\kms\ for MIKE, respectively. The grey band in the bottom panel shows the WiFeS primary rms residual.}
   \label{fig:rvs}
\end{figure}

\begin{figure*}
   \centering
   \includegraphics[width=\linewidth]{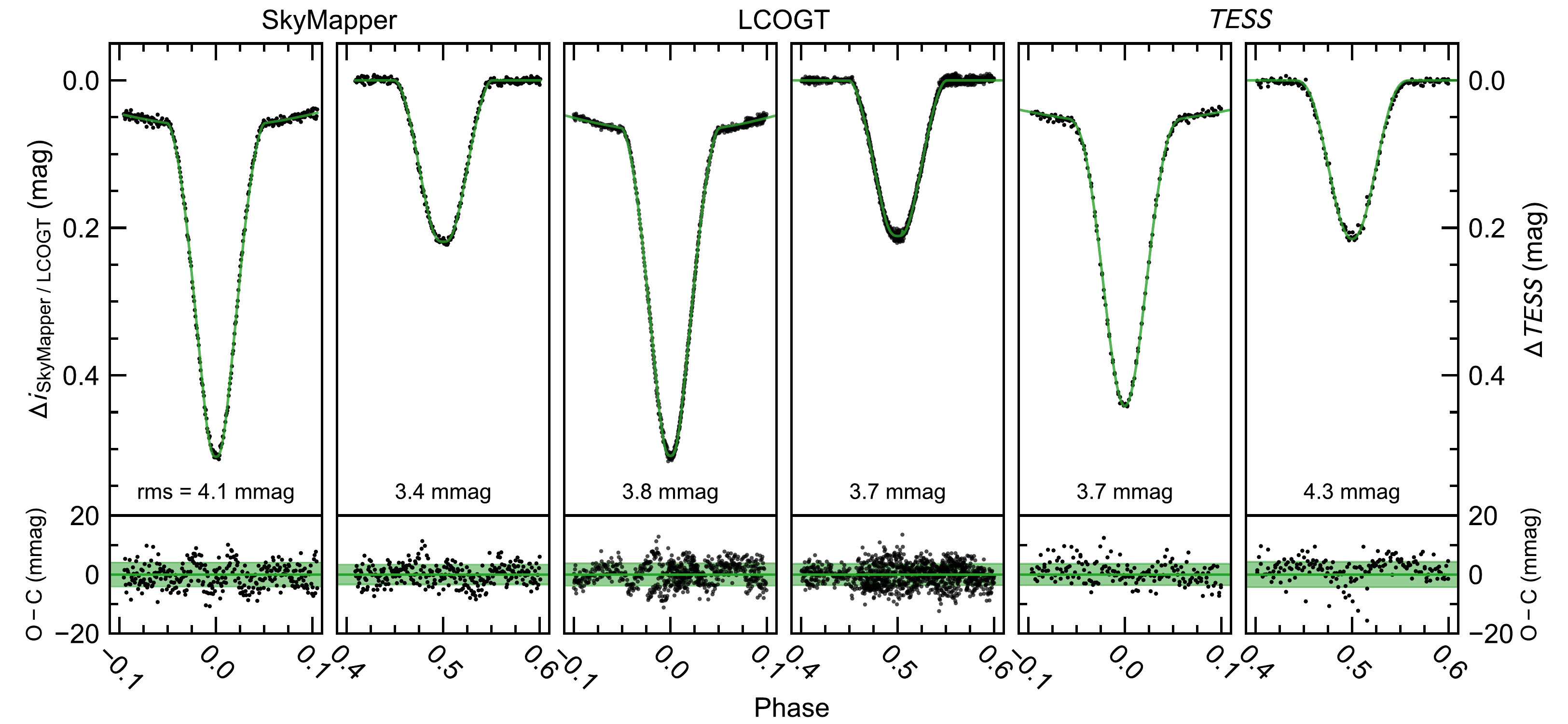}%
   \caption{Eclipses of \eb\ in each of the three photometric data sets used in this work, with the best-fitting model shown in green. Fit residuals are given in the bottom panel of each plot; the rms residual (green shaded bar) is calculated only for the ranges plotted here. The SkyMapper photometry cover only the eclipses and have been corrected with a quadratic zero-point in time, as described in Sec \ref{sec:lc_modelling}.}
   \label{fig:eclipses}
\end{figure*}

The final light curve and radial velocity solutions (corresponding to the median parameters in \autoref{table:parameters}) are plotted in Figs.\,\ref{fig:best_fit}, \ref{fig:rvs} and \ref{fig:eclipses}. The single-spot, spherical-star model is able to reproduce the observations with good fidelity, with rms residuals of 3--4 mmag and WiFeS (MIKE) velocity residuals of 3--4\,\kms\ (1--2\,\kms), in agreement with the input uncertainties and jitter terms. There is some structure visible in the light curve residuals, particularly around the eclipses (\autoref{fig:eclipses}). This is likely due to the adopted limb darkening coefficients, which are not well-constrained by the observations and tended to their maximal ($q_1$) or minimal ($q_2$) values in the fit. The structure in the LCOGT residuals is dominated by the effects of combining (uncorrected) light curve segments collected over many nights (\autoref{fig:lcogt}). 

The modelling yields masses of \ebmp\ and \ebms\,\Msun\ for the primary and secondary components, respectively, with radii of \ebrp\ and \ebrs\,\Rsun. These produce identical surface gravities of $\log g=\ebloggp$ and \ebloggs. The mass and radius uncertainties of $\sim$1~per cent and $\sim$0.5~per cent, respectively, are well within the 2~per cent convention suitable for the strictest tests of stellar models \citep{Andersen91,Southworth15}. The stars move on tight ($a=\ebarsun$\,\Rsun; \ebaau\,au) \ebperiod\,d orbits which are almost circular ($e=\ebeccnoerr$).  The ratio of stellar radius to effective Roche lobe radius is $\sim$0.4 for both stars \citep{Eggleton83}, confirming that the system is  detached and the stars are essentially spherical (see below).  

\begin{figure}
   \centering
   \includegraphics[width=\linewidth]{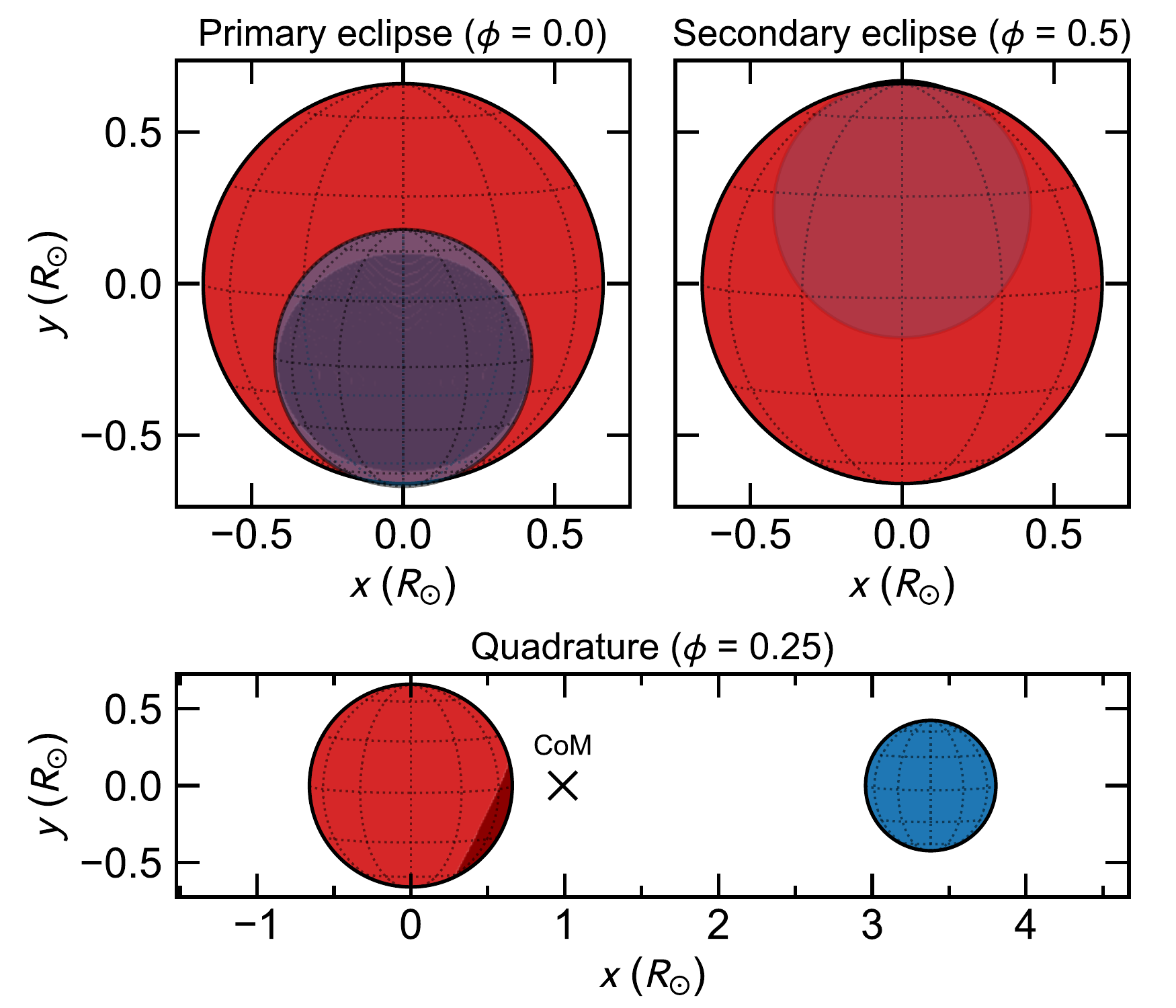}%
   \caption{Configuration of \eb\ as seen from Earth at primary eclipse, secondary eclipse and quadrature ($\phi=0.25$). The radii and separation of the components is drawn to scale, with the shaded region on the primary (red) showing the spot inferred in the fitting process. The secondary star (blue) is drawn partially transparent at $\phi=0$ to show the spot underneath. The centre of mass (CoM) of the system is marked.}
   \label{fig:configuration}
\end{figure}

The configuration of the system as seen from Earth is depicted in \autoref{fig:configuration}.  The secondary is almost entirely occulted at $\phi=0.5$, which is responsible for the deep secondary eclipse. Interestingly, the spot subtends approximately the same solid angle as the secondary and is fully covered by it at the midpoint of the primary eclipse. As the impact parameter of the eclipse, $b=a\cos(i)/ R_{1}$, depends on the orbital inclination (which is independent of the spot size or stellar radius) it seems unlikely that this configuration is coincidental and must therefore be a consequence of the simple spot model. A spot radius of 39$^{\circ}$ (covering $\sim$12 per cent of the stellar surface) is on the high end of inferred sizes reported in the literature for similar systems \citep[e.g.][]{Torres02,Ribas03,Lopez-Morales05}. A more realistic configuration would be smaller, spot complexes with larger brightness contrasts subtending the same approximate surface area as the single spot. Assuming the spot emits thermally, the $\sim$0.8~brightness ratio corresponds to a temperature ratio of $T_{\rm spot}/T_{\rm star} \approx 0.95$, in agreement with literature estimates. Further modelling is clearly required to better understand the true spot distribution on both stars. Finally, we note that even if the out-of-eclipse variation is not caused by spot modulation \citep[or is removed prior to fitting; e.g.][]{David19}, the radii derived from the light curve fit should not be significantly affected.

\subsubsection{Robustness of the results}

\begin{figure}
   \centering
   \includegraphics[width=\linewidth]{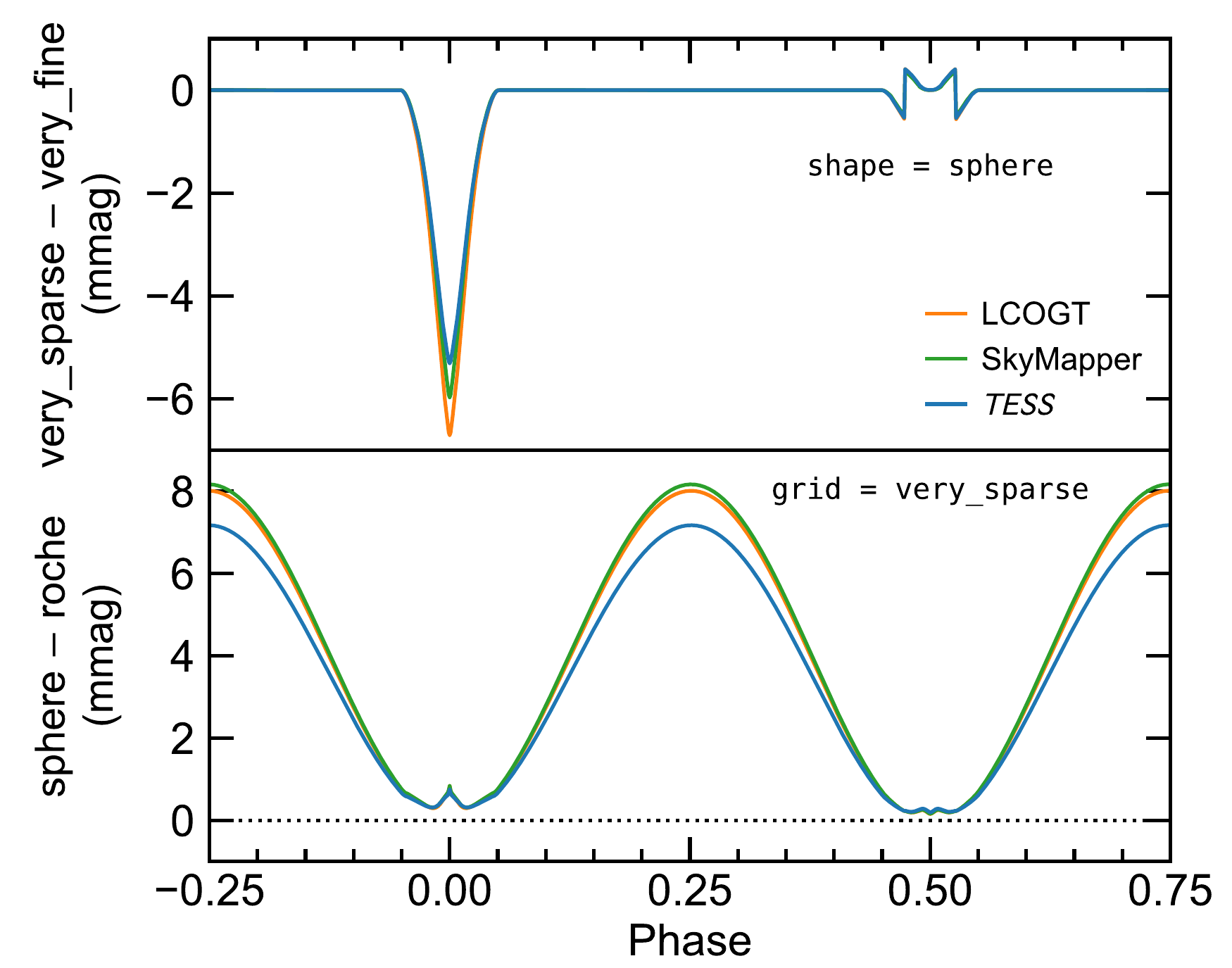}%
   \caption{Top panel: light curve difference between the adopted \texttt{very\_sparse} integration grid and the \texttt{very\_fine} grid in \ellc\ for each of the three photometric data sets. Bottom panel: difference between spherical stars and \ellc's Roche ellipsoids. In both panels the light curves were generated using the best-fitting parameters from \autoref{table:parameters}.}
   \label{fig:ellc}
\end{figure}

As noted above, to increase the speed of the MCMC analysis we used the sparsest integration grid in \ellc\ and assumed spherical stars. Here we test the effect these assumptions have on the system modelling. In \autoref{fig:ellc} we plot the difference between light curves generated from the parameters in \autoref{table:parameters} using the \texttt{very\_sparse} ($n=4$) and \texttt{very\_fine} ($n=32$) integration grids. The largest discrepancy is during the primary eclipse, with the \texttt{very\_fine} grid giving a 6--7~mmag deeper eclipse at $\phi=0$. By comparison, the fit residuals over the eclipse using the \texttt{very\_sparse} grid are 3--4 mmag (\autoref{fig:eclipses}). Given the large eclipse depth ($\sim$0.45 mag), adopting the sparse grid will therefore result in at most a 1--2~per cent systematic effect on the derived radii.

We also plot in \autoref{fig:ellc} the difference between light curves generated by spherical stars and those from triaxial ellipsoids assuming Roche geometry. As expected, the ellipsoidal variation is a half-period sinusoid with maxima at $\phi=0.25$ and 0.75. While it may change the inferred spot parameters, the low $\sim$8~mmag amplitude and minima around the eclipses means that adopting Roche geometry  should not significantly affect the radii estimates.

\subsection{Spectral energy distribution}

\begin{table}
\hfil%
\begin{minipage}{0.86\linewidth}
   \caption{Survey photometry of \eb.}
   \begin{tabular}{lccc} 
\hline
Bandpass & Magnitude & System & References \\
\hline
\gaia\ $G$ & $13.911\pm0.007$  & Vega  & (1,8) \\
\gaia\ $G_{\rm BP}$ & $15.45\pm0.03$  & Vega &(1,8) \\
\gaia\ $G_{\rm RP}$ & $12.69\pm0.02$  & Vega & (1,8)\\
SDSS $u$ & $18.82\pm0.02$  & AB &(2,9) \\
SDSS $g$ & $16.163\pm0.004$  & AB & (2,9)\\
SDSS $r$ & $14.793\pm0.004$  & AB &(2,9) \\
SDSS $i$ & $16.00\pm0.01^{a}$  & AB &(2,9) \\
SDSS $z$ & $12.550\pm0.004$  &AB &(2,9) \\
SkyMapper$^{b}$ $g$ & $15.51\pm0.01$  & AB &(3,10) \\
SkyMapper$^{b}$ $r$ &$14.56\pm0.01$   & AB &(3,10) \\
SkyMapper$^{b}$ $i$ &$12.963\pm0.005$   &AB & (3,10) \\
SkyMapper$^{b}$ $z$ &$12.344\pm0.004$  & AB &(3,10) \\
APASS $B$ & $16.7\pm0.1$  & Vega &(4,11) \\
APASS $V$ & $15.17\pm0.07$  & Vega & (4,11) \\
APASS $g$ & $15.8\pm0.1$  & AB&(4,11) \\
APASS $r$ & $14.50\pm0.08$  & AB & (4,11) \\
APASS $i$ & $13.17\pm0.09$  & AB &(4,11) \\
APASS $z$ & $12.49\pm0.02$  & AB &(4,11) \\
Pan-STARRS $g$ & $15.68\pm0.01$  & AB &(5,12) \\
Pan-STARRS $r$ & $14.49\pm0.01$  & AB &(5,12) \\
Pan-STARRS $i$ & $13.095^{c}$  & AB &(5,12) \\
Pan-STARRS $z$ & $12.545^{c}$  & AB &(5,12) \\
Pan-STARRS $y$ & $12.158\pm0.003$  & AB &(5,12) \\
2MASS $J$ & $10.93\pm0.03$  & Vega &(6,13)\\ 
2MASS $H$ &$10.29\pm0.02$  & Vega &(6,13)\\
2MASS $K_{s}$ & $10.04\pm0.02$  & Vega &(6,13)\\
All\emph{WISE} $W1$ & $9.96\pm0.02$  & Vega &(7,14)\\
All\emph{WISE} $W2$ & $9.80\pm0.02$  & Vega&(7,14)\\
All\emph{WISE} $W3$ & $9.71\pm0.05$   & Vega & (7,14)\\
\hline
   \end{tabular}\\
   $^{a}$Not included in SED fit. Significantly fainter than the SkyMapper, Pan-STARRS and APASS $i$-band fluxes.  SDSS  pipeline classifies the $i$-band detection as a Galaxy.\\
  $^{b}$SkyMapper DR2 photometry is currently only accessible to Australian researchers. The public DR1.1 release is available on VizieR \citep[II/358;][]{Wolf18}. \\
   $^{c}$Not included in SED fit. No uncertainty given in Pan-STARRS DR2 as all detections are saturated or too extended. \\
\emph{Photometry references:} (1) \emph\ \gaia\ DR2 \citep{Gaia-Collaboration18}; (2) SDSS DR15 \citep{Aguado19}; (3) SkyMapper DR2 \citep{Onken19}; (4) APASS DR10 \citep{Henden19}; (5) Pan-STARRS PS1 DR2 \citep{Chambers16}; (6)  \citet{Skrutskie06}; (7) \citet{Wright10}, \citet{Mainzer11}\\
\emph{Bandpass references:} (8) \citet{Evans18}; (9) SDSS DR7 website (\url{https://classic.sdss.org/dr7/instruments/imager/filters/index.html}); (10) \citet{Bessell11}; (11) \citet{Mann15}, SDSS; (12) \citet{Tonry12}; (13) \cite{Cohen03}; (14) \citet{Wright10} \\
  \label{table:photometry}
\end{minipage}
\hfil%
\end{table}

From the spectroscopic and light curve analysis we know that  \eb's spectral energy distribution (SED) is dominated by the M3.5 primary. To disentangle the contribution from the fainter secondary component we followed the method of \citet{Gillen17}, who analysed EBs in Praesepe and simultaneously fitted the component temperatures, reddening and distance by comparing model fluxes to optical and infrared photometry. 

We first gathered photometric observations for \eb\ from all available large surveys, noting if the photometry was absolutely calibrated to Vega or the AB magnitude system (or a combination of both, e.g. APASS).  We also acquired transmission curves for each bandpass and whether they were tabulated as photon$^{-1}$ or energy$^{-1}$ \citep[for further details see][]{Bessell12}. As no published bandpasses exist for the APASS survey \citep{Henden19}, we took Johnson $B$ and $V$ transmission curves from \citet{Mann15} and assumed the $griz$ bandpasses were the same as their SDSS equivalents.  These observations and references are given in \autoref{table:photometry}. We converted each observed magnitude to an in-band flux using either the latest CALSPEC \citep{Bohlin14}  Vega spectrum\footnote{CALSPEC name: \texttt{alpha\_lyr\_stis\_008}} or the $F_\nu = 3631$\,Jy AB reference spectrum \citep[converted to $F_\lambda$ through the bandpass pivot wavelength, see][]{Bessell12}.  We then computed synthetic fluxes for each bandpass using solar-metallicity BT-Settl \citep{Allard12,Baraffe15} and PHOENIX v2 \citep{Husser13} model atmospheres over a range of effective temperatures $2300\,\textrm{K} < T_{\rm eff} < 7000$\,K and surface gravities $2.5 < \log g < 5.5$. Before performing the synthetic photometry we linearly interpolated the spectra to 50\,K in $T_{\rm eff}$ and 0.25 in $\log g$.  The PHOENIX models only extend to 5\,\micron\  and so do not cover the \emph{WISE} $W3$ or $W4$ bandpasses\footnote{\eb\ has an All\emph{WISE} upper limit of $W4<8.793$ mag}.

During the fit the synthetic fluxes, $F(T_{\rm eff}, \log g)$, for each component were linearly interpolated from the model grid, diluted by the factor $(R/d)^2$ for radius $R$ and distance $d$, reddened to a given $E(B-V)$ and summed to give the combined flux. Note that the reddening vector  $A_{\lambda}$ is weakly dependent on the underlying stellar SED and magnitude of the extinction. Rather than redden the model spectra we calculated a representative $A_{\lambda}/E(B-V)$ for each bandpass using a 3300\,K, $\log g=4.5$ BT-Settl model with $E(B-V)=0.5$ and the  extinction law of \citet{Cardelli89}.  The free parameters in the fit were therefore the temperature, radius and mass of each component, the distance and the reddening. We also fitted for a photometric jitter term which was added in quadrature to the magnitude errors. This accounts for additional uncertainties not captured by the observations, such as errors in the absolute calibration, the unknown phase of the observations, the effects of spots, etc.  

We imposed Gaussian priors on the masses and radii from \autoref{table:parameters} and adopted a \gaia\ DR2 distance prior of $102.9\pm0.6$~pc \citep{Bailer-Jones18}.  With accurate component masses, radii and a distance, the surface gravities and dilution factors are essentially fixed, leaving only the temperatures and reddening to be determined. We adopted uniform priors on the temperatures and a Gaussian prior of $E(B-V)=0.03\pm0.01$~mag on the reddening towards the 32~Ori Group from \citetalias{Bell17}. This minimal reddening agrees with recent 3D reddening maps \citep[e.g.][]{Lallement18}. We also imposed a prior on the $i$-band surface brightness ratio, $J_{\rm LCOGT} = F_2/F_1 =\ebsb$ which acts as a temperature constraint but is directly related to the observed light curve. We explored the posterior parameter space of the model using \textsc{emcee} with 128 walkers and 10\,000 steps, conservatively retaining the last half of each chain for parameter estimation. 

\begin{table}
\hfil%
\begin{minipage}{0.89\linewidth}
   \caption{Results of the SED fitting against BT-Settl and PHOENIX model atmospheres.  Component masses and radii are not listed as we essentially recover the priors from the joint modelling.}
   \begin{tabular}{lcc} 
\hline
Parameter & BT-Settl & PHOENIX \\
\hline
Primary $T_{\rm eff}$ (K) &  $3309\pm22$ & $3284\pm16$ \\
Secondary $T_{\rm eff}$ (K) & $3022\pm 13$ & $2992\pm12$ \\
$E(B-V)$ (mag) & $0.035\pm0.01$ &  $0.035\pm0.01$ \\
Photometric jitter (mag) & $0.19\pm0.03$ & $0.15\pm0.03$\\[2mm]
Distance (pc; \gaia\ prior) & $102.8\pm0.6$ & $102.7\pm0.6$\\
Distance (pc; uniform prior)$^a$ & $100\pm5$  & $95\pm3$  \\[2mm]
Adopted primary $T_{\rm eff}$ (K)  & \multicolumn{2}{c}{$3293 \pm 13 \textrm{ (stat.)} \pm20$ (sys.)} \\
Adopted secondary $T_{\rm eff}$ (K) & \multicolumn{2}{c}{$3006 \pm 9 \textrm{ (stat.)} \pm20$ (sys.)} \\
Primary $\log(L/\Lsun)$ & \multicolumn{2}{c}{$-1.337\pm0.013$} \\
Secondary $\log(L/\Lsun)$  & \multicolumn{2}{c}{$-1.879\pm0.013$}\\
\hline
   \end{tabular}\\
   $^{a}$The BT-Settl and PHOENIX temperatures are approximately 20\,K and 50\,K cooler, respectively, than the \gaia\ prior.  \label{table:sed}
\end{minipage}
\hfil%
\end{table}

\begin{figure}
   \centering
   \includegraphics[width=\linewidth]{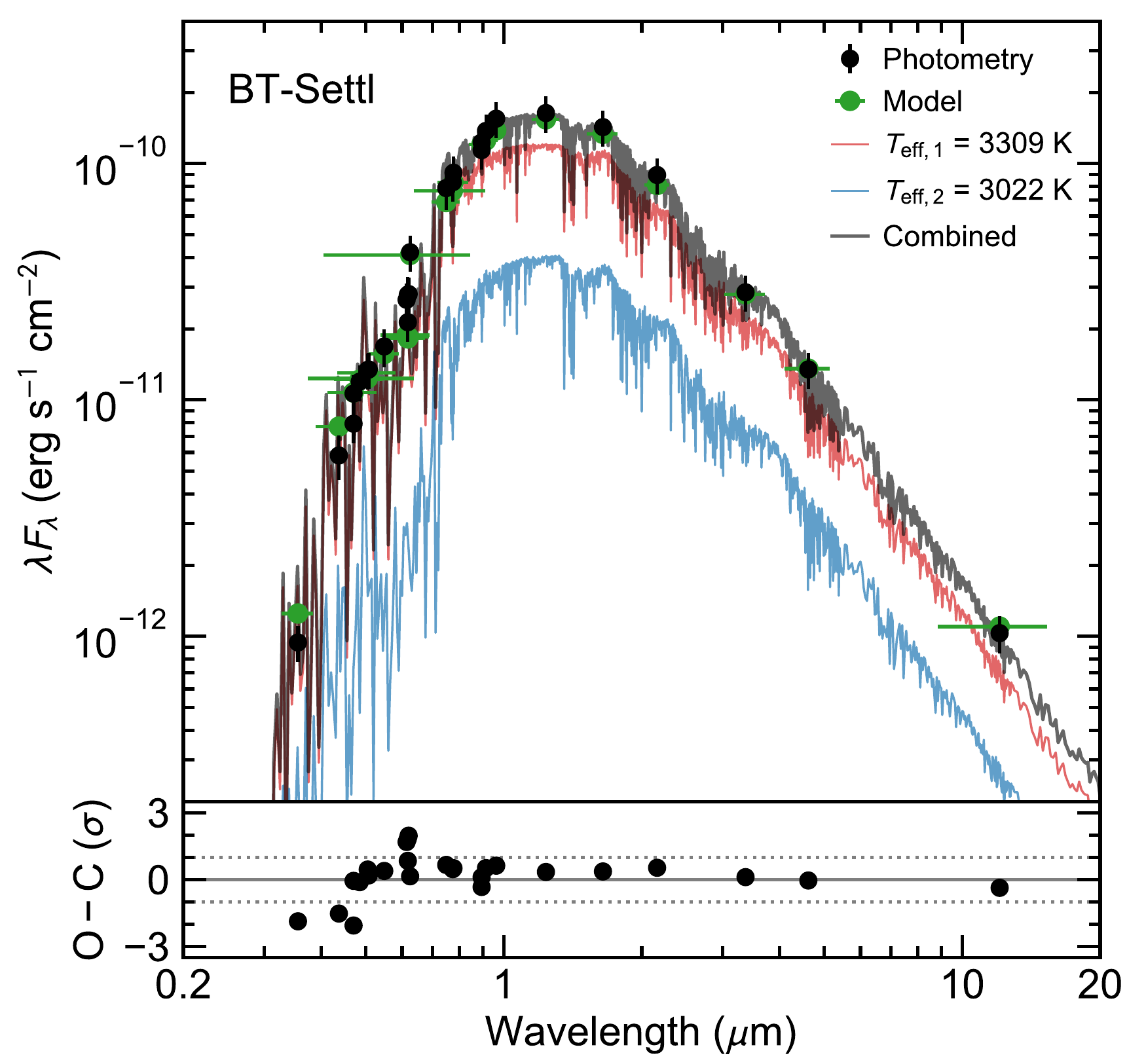}%
   \caption{Spectral energy distribution (SED) of \eb\ constructed from the magnitudes in \autoref{table:photometry} (black points). The best-fitting BT-Settl model atmospheres for the primary and secondary stars are plotted in red and blue, respectively. Their combined SED is given by the grey line and green points. Horizontal error bars show the equivalent width of each bandpass.}
   \label{fig:sed}
\end{figure}

The results of these fits are presented in \autoref{table:sed} and the SED of \eb\ is plotted against  BT-Settl model fluxes in \autoref{fig:sed}. Both sets of model atmospheres are able to reproduce the observed SED of \eb\ with a median reddening slightly larger than the prior. As expected, we recover our tight priors on the masses, radii, surface brightness ratio and distance. We note that the photometric jitter term completely dominates the observed photometric errors in both cases. Like \citet{Gillen17}, we find that the BT-Settl models under-predict the $r$-band fluxes and give slightly ($\sim$30\,K) warmer temperatures than the PHOENIX models, although they agree to within 2$\sigma$. Surprisingly, both sets of models over-predict the SDSS $u$-band flux, where we might expect a young system like \eb\ to have enhanced blue emission from its active chromosphere. It is therefore possible that this measurement is erroneous.\footnote{SkyMapper DR1.1 \citep{Wolf18} includes a brighter $u$-band magnitude of $17.97\pm0.26$ for \eb\ from two observations, which is near the faint limit of its Shallow Survey. However, none of these or subsequent observations were of high enough quality for inclusion in DR2.} The other two discrepant blue points in \autoref{fig:sed} are the APASS $B$ and SDSS $g$-band fluxes. Neither is a significant outlier when using the PHOENIX models, but we note that the SDSS $g$-band magnitude is 0.3--0.6 mag fainter than the SkyMapper, Pan-STARRS or APASS measurements (see \autoref{table:photometry}).

We adopt the weighted mean of the BT-Settl and PHOENIX models as our final temperature; this is \ebteffp\,K for the primary and \ebteffs\,K for the secondary. The corresponding luminosities from the Stefan--Boltzmann law are $\log(L/\Lsun) = \ebloglp$ and \eblogls, where we have included an additional 20 K systematic error on each temperature and used the updated solar constants from \citet{Mamajek15b}.  The primary temperature agrees with the 3300\,K expected of a M3.5 star from the pre-main sequence temperature scale of \citet{Herczeg14}, and is bracketed between the \citet{Luhman03}  and \citet{Pecaut13b} predictions of 3340\,K and 3260\,K, respectively. The \citeauthor{Luhman03} scale is more appropriate for few Myr-old stars with intermediate gravities between dwarfs and giants and so may not be suitable for a system as old as \eb.

As a test we also re-fitted with a uniform distance prior, this gave median distances of \ebdistbtsettl\ pc and \ebdistphoenix\ pc for the BT-Settl and PHOENIX models, respectively, with temperatures $\sim$20\,K and $\sim$50\,K cooler than the \gaia\ prior. The good agreement between these distances and the 102.9~pc from \gaia\  shows that the radii inferred from the light curve modelling are accurate. A uniform prior on $E(B-V)$ produced median reddenings of $\sim$0.2~mag and temperatures 100--200\,K warmer than the \citetalias{Bell17} prior. Such a high reddening is not supported by the 3D maps or the close agreement between the spectra of \eb\ and GJ 273 (\autoref{fig:wifes_3000}), which at a distance of 3.8~pc is essentially unreddened.

\subsection{Rotation and activity}
\label{sec:rotation}

As outlined above, the short period of \eb\  implies that the two components should be rotating synchronously and have circular orbits.  Using the period (\ebperiod\,d) and mass ratio ($q=0.41$) derived from the joint modelling, we estimate an approximate synchronization time-scale of $\sim$16 kyr and a circularization time-scale of 0.6 Myr \citep[][equations 6.1 and 6.2]{Zahn77}.  Clearly these are both much less than the $24\pm4$ Myr age of the 32 Ori Group \citepalias{Bell17} which we ascribe to \eb.  

\begin{figure} 
   \centering
   \includegraphics[width=\linewidth]{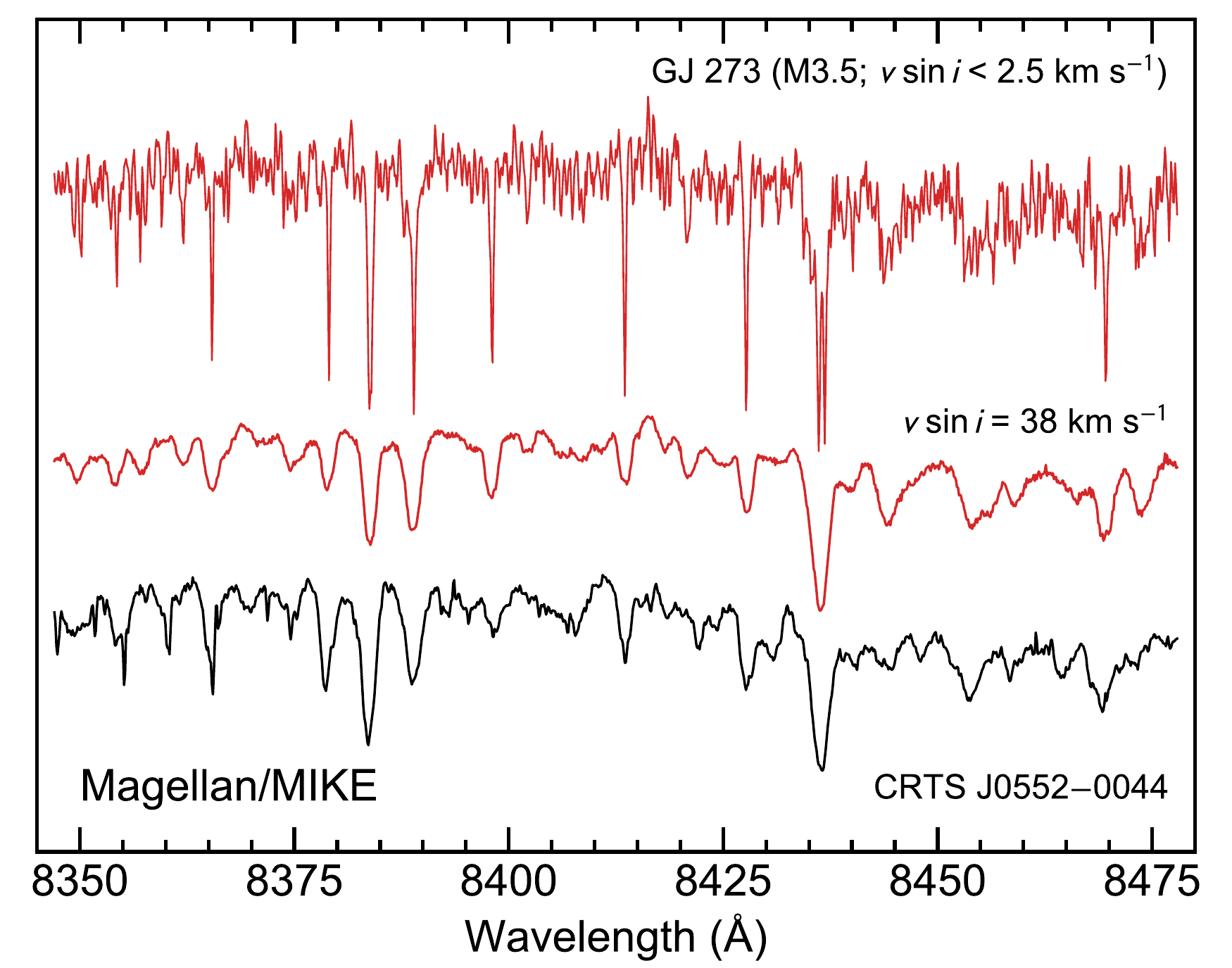}%
   \caption{Magellan/MIKE spectrum of \eb\ (bottom) compared to the M3.5 slow-rotator GJ 273 (top). By broadening the latter to $\vsini \approx 38$\,\kms\ we can reproduce the broad lines observed in \eb. Note the actual determination of \vsini\ was made using the cross-correlation function width and not by comparing spectra directly.}
   \label{fig:vsini}
\end{figure}

Circularization of the orbits appears to have taken place -- our best-fitting eccentricity ($e=\ebecc$) is consistent with zero, and assuming synchronous rotation we calculate a projected rotation rate for the primary of $\vsini_{\rm calc} = \ebvsinicalc$\,\kms. The uncertainty reflects the variation in the radius, period and orbital inclination from \autoref{table:parameters}. As discussed in \autoref{sec:magellan}, from the 12 Magellan/MIKE measurements we find a weighted mean of $\vsini_{\rm obs} =  \ebvsiniobs$\,\kms, in excellent agreement with the synchronous estimate. Assuming the out-of-eclipse variability is dominated by spots, the stability of the spot pattern in phase over the $\sim$1~yr baseline of observations presented here ($>$400 orbits) is further evidence of synchronicity.  The effect of this fast rotation is readily apparent in \autoref{fig:vsini}, where we compare a portion of the MIKE spectrum of \eb\ to the slow-rotating template GJ 273 ($\vsini < 2.5$\,\kms). By broadening the latter we find a best-fitting $\vsini\ \approx 38$\,\kms\ for this region, in agreement with the value derived from the CCF width calibration.

This fast rotation gives rise to enhanced magnetic activity, manifest in \eb\ by strong \halpha\ and other Balmer line emission, \ion{Ca}{ii} H\,\&\,K emission and the filled-in absorption lines of \ion{Na}{i} D and the \ion{Ca}{ii} IRT (\autoref{fig:wifes_3000}). It should also produce strong coronal X-ray emission.  \emph{ROSAT} detected a point source, 2RXS J055257.6$-$004422, $\sim$30 arcsec to the east of \eb\  with a positional error of 19 arcsec (see \autoref{fig:tess_tpf}). Assuming this source is associated with \eb\footnote{In their automated catalogue of radio/X-ray associations to optical sources (MORX), \citet{Flesch16} associated 2RXS J055257.6$-$004422 with a faint ($R\sim19.2$ mag) optical source 16 arcsec south-west of the \emph{ROSAT} detection. They note, however, that there is a 56 per cent probability the association is false. Given the positional error on the detection and expectation of copious X-ray emission from a fast rotating young binary, we believe 2RXS J055257.6$-$004422 is more likely associated with \eb.}, we use the count rate and hardness ratio from the Second \emph{ROSAT} All-Sky Survey \citep[2RXS;][]{Boller16} and the conversion factor of \citet{Fleming95b} to derive a 0.1--2.4~keV flux of $\sim$$3\times10^{-13}$~erg~cm$^{-2}$~s$^{-1}$.  We note that this detection is near the \emph{ROSAT} faint limit, with errors on the count rate and hardness ratio of $\sim$30 per cent and $\sim$100 per cent, respectively. With a system luminosity of $\sim$0.06\,\Lsun, we find an X-ray to bolometric luminosity ratio of $\log(L_{\rm X}/L_{\rm bol}) \sim -2.8$, consistent with the $\log(L_{\rm X}/L_{\rm bol})\approx -3$ saturated X-ray emission seen in young and active stars \citep[e.g.][]{Fleming89,Stauffer97}. Finally, we note that the 2RXS binned light curve ($\Delta t = \textrm{3.2}$\,h) is consistent with a constant count rate over the $\sim$44\,h baseline of the \emph{ROSAT} observations, with no flares apparent. While a quantitative discussion of flare rate and strength in \eb\ is beyond the scope of this work, the detection of several low-level flares in the LCOGT and \tess\ light curves (e.g. \autoref{fig:tess_lc}) is consistent with the enhanced activity observed in young, low-mass stars and close binaries.

\vspace{-2mm}
\section{Discussion}
\label{sec:discussion}

\subsection{\eb\ in context}

\begin{figure*}
   \centering
   \includegraphics[width=\linewidth]{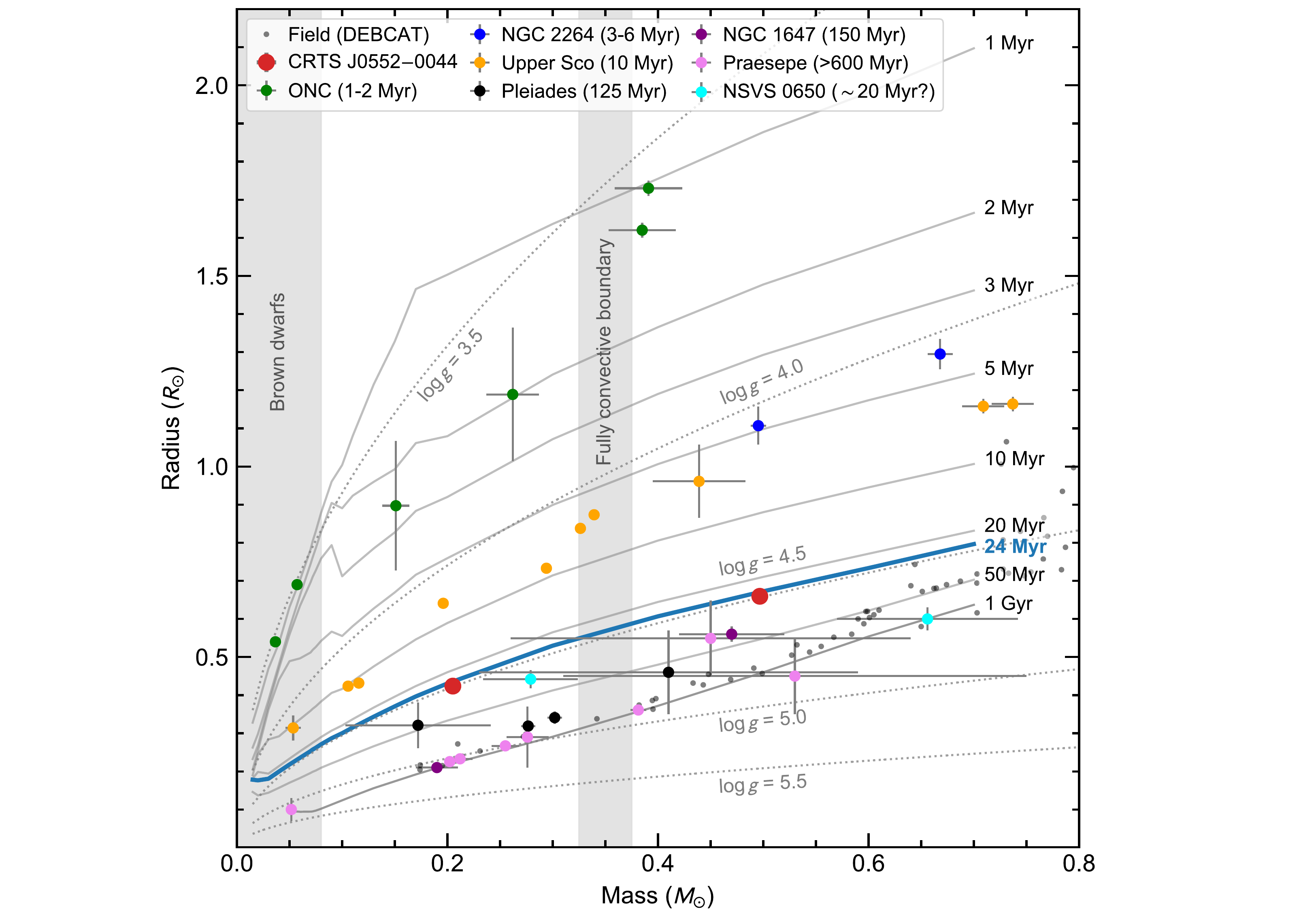}%
   \caption{Mass--radius diagram comparing \eb\ (large red points; error bars smaller than the point size) to well-characterized field-age EBs \citep[DEBCAT;][]{Southworth15} and young ($<$1 Gyr), low-mass systems collated by \citet{Gillen17} and \citet{David19}. Note that the Upper Sco star with the largest errors is a single-lined system (RIK 72) with an inferred $\sim$0.05\,\Msun\ companion (also plotted). We also include the components of NSVS 06507557 \citep{Cakirli10}, which is claimed to have an age of $\sim$20 Myr (see text). Overplotted are solar metallicity \citet{Baraffe15} isochrones (solid lines) and surface gravity contours from $\log g=3.5$ to 5.5 (dotted lines). The thick blue line is the 24 Myr isochrone at the age of the 32 Ori Group and the shaded region indicates the boundary between partially and fully-convective ($\lesssim$0.35\,\Msun) stellar interiors on the main sequence.}
   \label{fig:radius_mass}
\end{figure*}

We plot in \autoref{fig:radius_mass} our radius and mass measurements compared to the pre-main sequence evolutionary models of \citet{Baraffe15} (hereafter \citetalias{Baraffe15}). The components of \eb\ are coeval in the mass--radius diagram and have larger radii and lower surface gravities ($\log g=4.5$) than older field stars of the same mass, as expected of young stars finishing their contraction onto the main sequence.  Note that the $\log g=4.5$ line of constant surface gravity is essentially parallel to the models at these ages. Both components appear slightly older than the 24~Myr BHAC15 isochrone at the canonical age of the 32~Ori Group. After creating a denser set of isochrones (\mbox{$0<\log t/\textrm{Myr} < 2$}; $\Delta\log t = 0.02$), a linear interpolation\footnote{Using  \texttt{scipy.interpolate.LinearNDInterpolator} in \textsc{python}.} of the models yields ages of 25.4 and 26.2~Myr for the primary and secondary, respectively. We compare the full gamut of system properties (masses, radii, surface gravities, temperatures, luminosities) to the BHAC15 and other contemporary pre-main sequence models in \autoref{sec:models}. 

Also plotted in \autoref{fig:radius_mass} is the census of young ($<$1 Gyr), low-mass ($<$0.8\,\Msun) EBs collated by \citet{Gillen17} and \citet{David19}. This ensemble provides the best-available constraints on  evolutionary models at the lowest masses and ages \citep[e.g.][]{Stassun14}.  \eb\ is one of only 17 such systems identified to date and only the second low-mass EB at intermediate ages between young groups like the ONC (1--2~Myr) and Upper Scorpius (5--10~Myr), and older populations like the Pleiades (125~Myr) and Praesepe ($\sim$600~Myr). The other system is NSVS 06507557 \citep{Cakirli10} which is claimed to have an age of $\sim$20~Myr.  However, its component ages inferred from our mass--radius diagram are inconclusive. The primary lies on the 1~Gyr  isochrone while the secondary is about 35~Myr old, with a radius $\sim$50 per cent larger than its expected main sequence value. The fast-rotating  ($P=0.52$\,d) system shows evidence of spot activity and its optical spectrum has \ion{Li}{i} absorption as well as Balmer and forbidden line emission characteristic of young stars.  It is not a known member of any nearby star-forming region or moving group. There is a faint star 4~arcsec to the south with a nearly identical \emph{Gaia} parallax and proper motion, so it is possible this wide companion has complicated the light curve or velocity analysis, or is responsible for the spectroscopic youth indicators. Regardless of its exact age, the masses derived by \citet{Cakirli10} are not precise enough for stringent testing of evolutionary models.  

At higher masses ($>$0.8\,\Msun), \eb\ joins the intermediate age systems NP Per \citep[$1.3+1.0$\,\Msun, $\sim$17 Myr;][]{Lacy16}, MML 53 \citep[$1.0+0.9+0.7$\,\Msun, 16 Myr;][]{Hebb10,Hebb11,Gomez-Maqueo-Chew19} and several solar-type EBs in the greater Orion OB1 association \citep{Stassun14}. MML 53 in particular has a well-constrained age from its membership in the Upper Centaurus Lupus subgroup of the Sco-Cen OB association \citep[$16\pm2$~Myr;][]{Pecaut16}. 

Low-mass EBs with mass ratios $\ll$1 are especially valuable for testing evolutionary models as they allow an  assessment of the model predictions over a wide range of masses at a single age and metallicity.  Among the young systems in \autoref{fig:radius_mass}, \eb\ has one of smallest mass ratios ($q=0.41$) and is one of only six systems spanning the $\sim$0.35\,\Msun\ main sequence fully-convective boundary \citep{Chabrier97}. The others are NSVS 06507557, RIK 72 \citep[single-lined system;][]{David19}, and older members of the Pleiades \citep[MHO~9;][]{David16a}, Praesepe \citep[AD~3814;][]{Gillen17} and NGC 1647 \citep[2MJ0446$+$19;][]{Hebb06}.  \eb\ could be considered a  younger analogue of 2MJ0446$+$19, which has a similar orbital period (0.62 d), component masses ($0.47+0.19$\,\Msun; $q\sim0.4$) and temperatures (3320~K, 2910~K). In contrast to NSVS 06507557, the primary component of 2MJ0446$+$19 (age $\sim$150 Myr) is significantly inflated while the secondary lies on the 1~Gyr isochrone. 

\begin{figure}
   \centering
   \includegraphics[width=\linewidth]{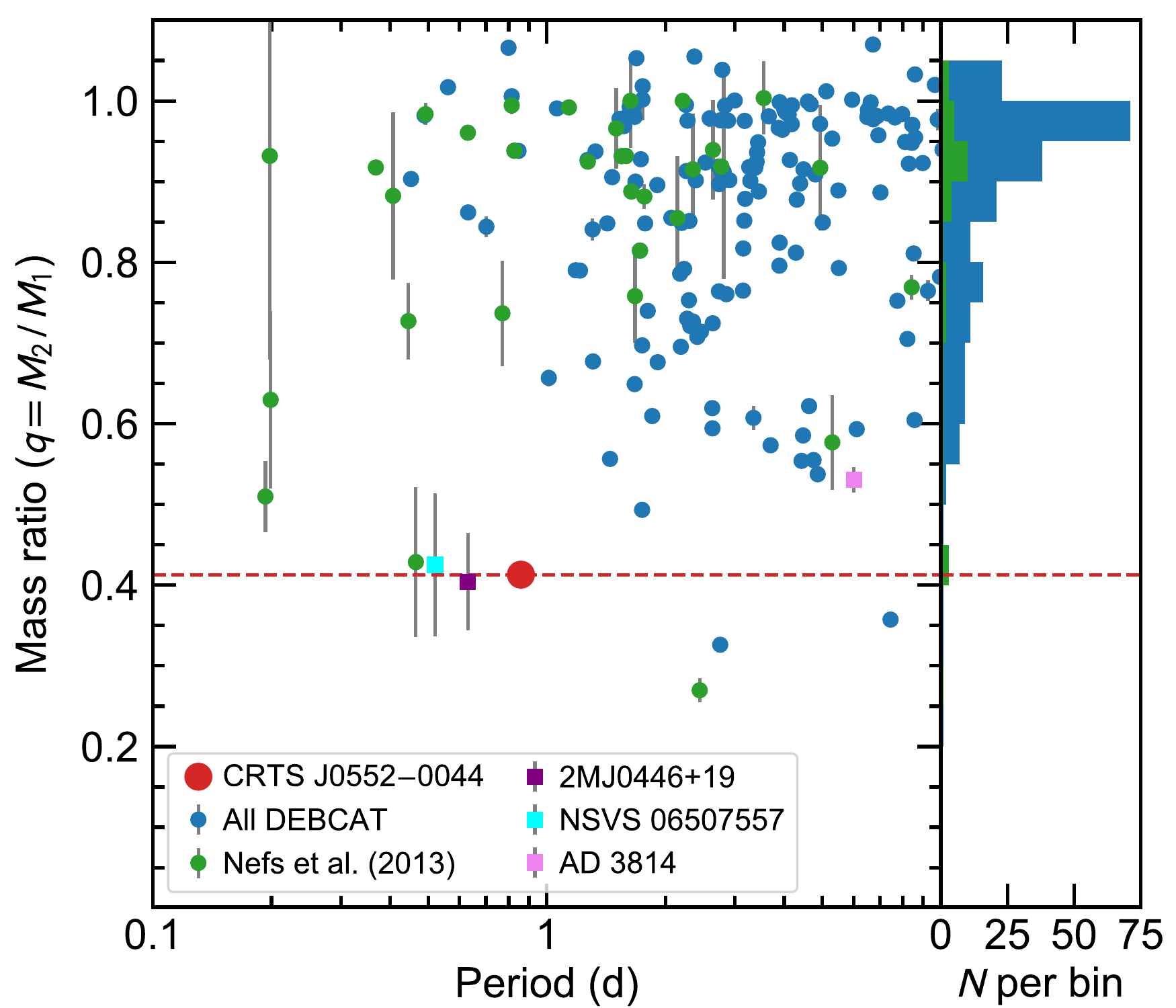}%
   \caption{Mass ratio and period distribution of EBs from DEBCAT \citep[blue points;][]{Southworth15} and short-period ($P<10$~d) M dwarf systems from \citet{Nefs13} (green points). The three short-period systems from \autoref{fig:radius_mass} which span the fully-convective boundary are plotted as coloured squares. The histogram gives the mass ratio distribution for the \citeauthor{Nefs13} stars and the full DEBCAT sample.}
   \label{fig:mass_ratio}
\end{figure}

Star formation theories predict that short-period M dwarf binaries with low mass ratios should be rare \citep{Nefs13}. This is due to primordial factors, such as the nascent secondary preferentially accreting infalling gas with high angular momentum, driving the mass ratio to unity \citep[e.g.][]{Bate00,Bate02}, as well as dynamical effects in the young cluster. Both the dynamical decay of small-$N$ multiples or exchanges with single stars will tend to tighten orbits and eject the lowest-mass components \citep[e.g.][]{Sterzik95,Goodwin07}. The frequency of short-period,  unequal-mass binaries is therefore expected to decrease steeply with the mass of the primary.  

The present-day result of these processes is illustrated in \autoref{fig:mass_ratio}, where we compare \eb\ in period--mass ratio space to the high-precision (mass and radius errors $\lesssim$2~per cent) \mbox{DEBCAT}\footnote{\url{https://www.astro.keele.ac.uk/jkt/debcat/}} sample \citep{Southworth15} and short-period ($P<10$ d) M~dwarf binaries collated by \citet{Nefs13}. While the majority of systems have mass ratios close to unity, \eb\ falls in a region of the diagram not well populated by either sample. Interestingly, its position is very similar to the young EBs 2MJ0446$+$19 and NSVS 06507557 (see above).  Given the dynamical processing that occurs during pre-main sequence evolution, we expect the youngest systems to have lower mass ratios. However, in all three cases the EBs are found in older ($>$20 Myr), sparser groups or in the field. While it is a small sample, we may be seeing the effect of an isolated environment on binary properties. \eb\ and the other systems may have formed in isolation, been ejected from their natal groups or drifted out when the groups lost their binding gas. In any case, \eb\ was not further disrupted and it survives to the present day on the outskirts of the 32 Ori Group. 

Finally, as noted by \citet{Nefs13}, there is almost certainly an observational bias towards equal-mass M+M binaries because of the steep mass--luminosity relationship for M dwarfs.  Indeed, in the case of \eb\ we only detected the secondary through the velocity shift of its \halpha\ emission line, which was visible because of the enhanced chromospheric activity prevalent at lower masses. 

\subsection{Membership in the 32 Orionis Moving Group}

\begin{figure*}
   \centering
   \includegraphics[width=\textwidth]{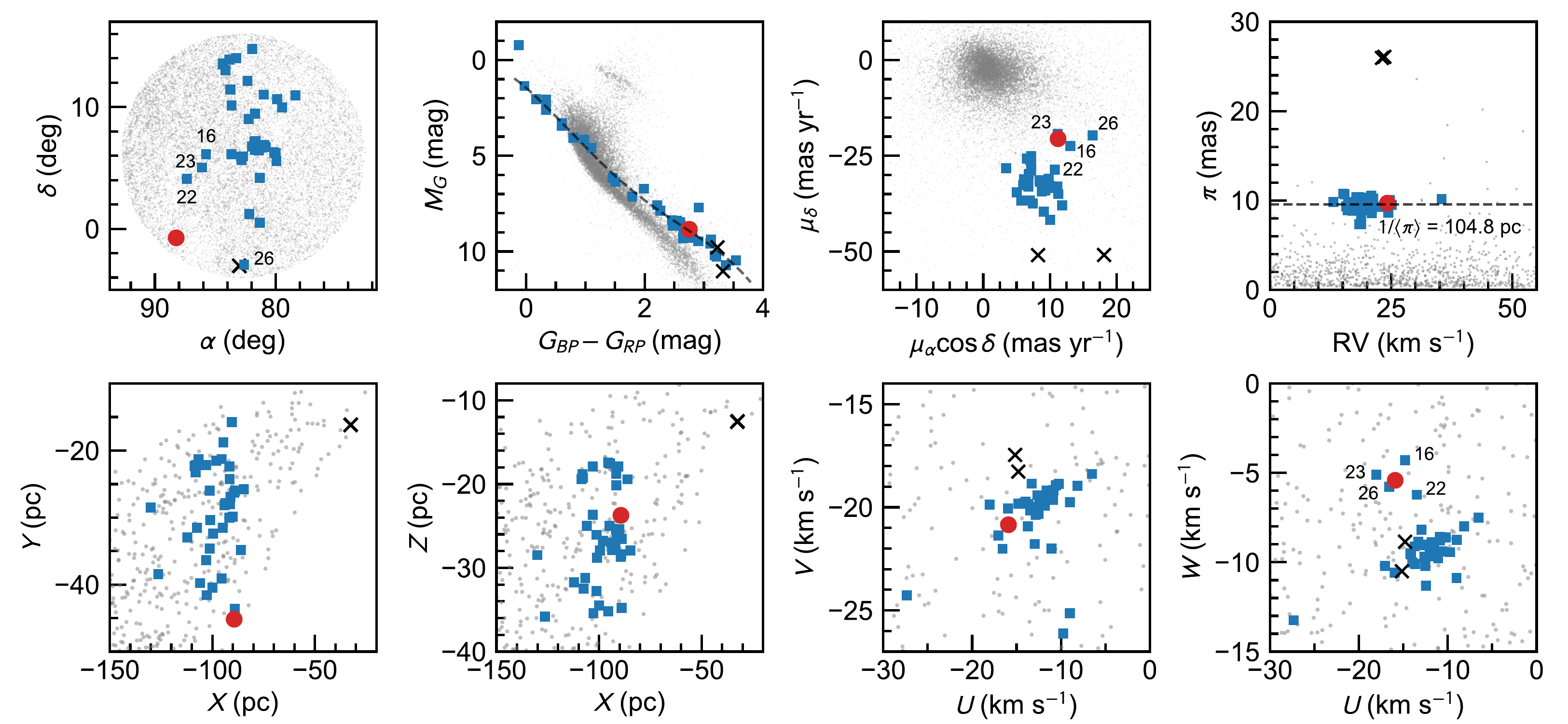}%
   \caption{Top row: Proposed members of the 32 Ori Moving Group from \citetalias{Bell17} (blue squares) compared to \eb\ (red circle) in \gaia\ DR2 observables (left to right): sky position, photometry, proper motion and parallax (with radial velocities from \citetalias{Bell17}). The two new non-members THOR 33 and 34 are shown as black crosses. Grey points are a random sample of 20\,000 DR2 sources within 10$^{\circ}$ of 32 Ori itself with $\upi/\sigma_{\upi}>10$. Only THOR 18 (SCR 05220606) is outside this 10$^{\circ}$ limit.  A quartic polynomial has been fitted to members in the CMD to guide the eye. Bottom row: the same observables transformed to heliocentric $XYZ$ positions and $UVW$ velocities. Members mentioned in the text are labelled by their THOR number from \citetalias{Bell17}.}
   \label{fig:xyzuvw}
\end{figure*}

Before a thorough comparison to evolutionary models, it is prudent to re-examine the proposed membership of \eb\ in the 32~Ori Group.  \citetalias{Bell17} assigned \eb\ as a possible member  based on its UCAC4 proper motion, 92~pc kinematic distance, colour--magnitude diagram (CMD) position and preliminary systemic velocity ($20.9\pm2.3$\,\kms). With an improved velocity and high-precision parallaxes and proper motions from the \gaia\ mission \citep{Gaia-Collaboration18}, we can more confidently claim membership in the group.

Of the 46 group members proposed in \citetalias{Bell17}, only the L1 brown dwarf THOR 41 \citep[=WISE J052857.69+090104.2; ][]{Burgasser16} is not found in \gaia\ DR2.  All but five members have parallaxes and proper motions. We plot the DR2 observables  and heliocentric position ($XYZ$) and velocity ($UVW$) vectors for these stars and \eb\ in \autoref{fig:xyzuvw}. While a full discussion of every group member in light of \gaia\ astrometry is outside the scope of this work, we can immediately reject THOR 33AB (5~arcsec binary) and THOR 34 (3~arcmin from 33AB) as members due to their significantly larger parallaxes (distance $\sim$38\,pc).  Their position and space motion make them probable members of the similarly-aged $\beta$~Pictoris Moving Group (BPMG).\footnote{Using \gaia\ astrometry and radial velocities from \citetalias{Bell17},  the \textsc{banyan}-$\Sigma$ tool \citep{Gagne18} returns membership probabilities of $>$99 per cent for the BPMG, with the balance of probabilities going to the young field.} The remaining 32 Ori Group members in \autoref{fig:xyzuvw} form a coherent association in proper motion, parallax, radial velocity and colour--magnitude space, confirming their status as a true moving~group. 

\eb\ lies on the south-east periphery of the known membership, with a sky position, proper motion and $W$ velocity component similar to the  members THOR 16, 22, 23 and 26. Its DR2 distance is $<$2~pc from the group mean of 104.8~pc and its  systemic velocity places it only 4.8\,\kms\ (1.3$\sigma$) from the group mean of $19.5 \pm 3.7$\,\kms\ ($\pm$1~s.d.; after removing the A3 outlier HD 35714, whose velocity is likely unreliable). Unsurprisingly, the system does not lie above the locus of (single) 32 Ori Group members in the CMD, as its SED is dominated by the M3.5 primary. Based on the distribution of  members in \autoref{fig:xyzuvw} and radii consistent with $\sim$24~Myr isochrones,  we conclude that \eb\ is a bona fide member of the 32~Ori Group. Following \citetalias{Bell17}, we assign it the incremental membership number THOR~42.\footnote{\eb\ was classified by \citetalias{Bell17} as a \emph{possible} member and so not assigned a membership number in that study. As defined in \citetalias{Bell17} and adopted in recent works (e.g. \textsc{banyan}), the abbreviation THOR (= \emph{TH}irty-two \emph{OR}i) should not be confused with the distinct and much older Tucana--Horologium Association (often shortened to Tuc-Hor or THA).}

\begin{figure}
   \centering
   \includegraphics[width=\linewidth]{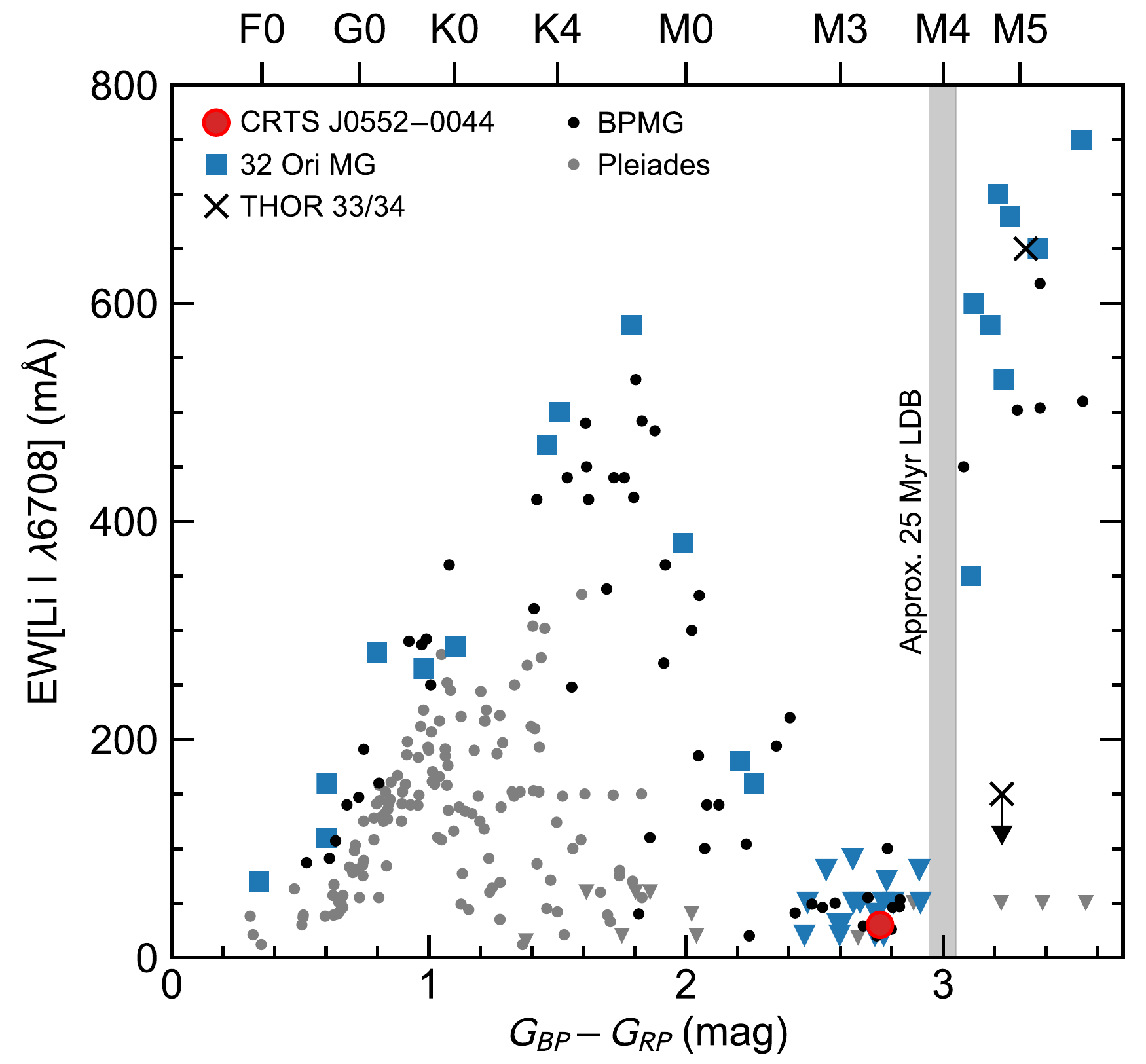}%
   \caption{\ion{Li}{i} $\lambda6708$ equivalent widths for members of the 32 Ori Group, the $\beta$ Pic Moving Group (BPMG) and the Pleiades. Upper limits are denoted by arrows. We detected no \ion{Li}{i} in \eb\ ($\mathrm{EW} \leq 30$~m\AA). The approximate location of the 25 Myr Lithium Depletion Boundary (LBD) observed in the 32 Ori Group and BPMG is given by the shaded region. }
   \label{fig:lithium}
\end{figure}

In their recent spectroscopic survey of the Orion OB1 association, \citet{Briceno19} reported \ion{Li}{i} $\lambda6708$ in \eb, with an equivalent width (EW) of 123~m\AA. This was just above their estimated $\sim$100\,m\AA\ detection limit. However, we did not detect \ion{Li}{i} $\lambda6708$  in any of our WiFeS or Magellan  spectra ($\mathrm{EW} \leq 30$~m\AA) and the EW distribution of 32 Ori Group members (\autoref{fig:lithium}) implies that an M3.5 star should not have detectable lithium at this age. Pre-main sequence evolutionary models predict that a 0.5\,\Msun\ star should have fully depleted its surface lithium in $\sim$10~Myr. The non-detection of lithium in \eb\ is therefore necessary but insufficient evidence of membership in the group.  

The mass (and hence luminosity) at which lithium remains unburnt in a stellar population is a sensitive function of age,  which has been exploited in recent years to precisely age-date several groups and open clusters younger than $\lesssim$200~Myr, including the 32~Ori Group. \citetalias{Bell17} found a mean lithium depletion boundary (LDB, see \autoref{fig:lithium}) age of $23\pm4$~Myr, which relied on kinematic distances with errors of 8--30 per cent. This agreed with the isochronal age of $25\pm5$~Myr derived using the same distances, giving the final age for the group of $24\pm4$~Myr, which we adopt here.

\subsection{Comparison to stellar evolutionary models}
\label{sec:models}

\begin{figure*}
   \centering
   \includegraphics[width=\linewidth]{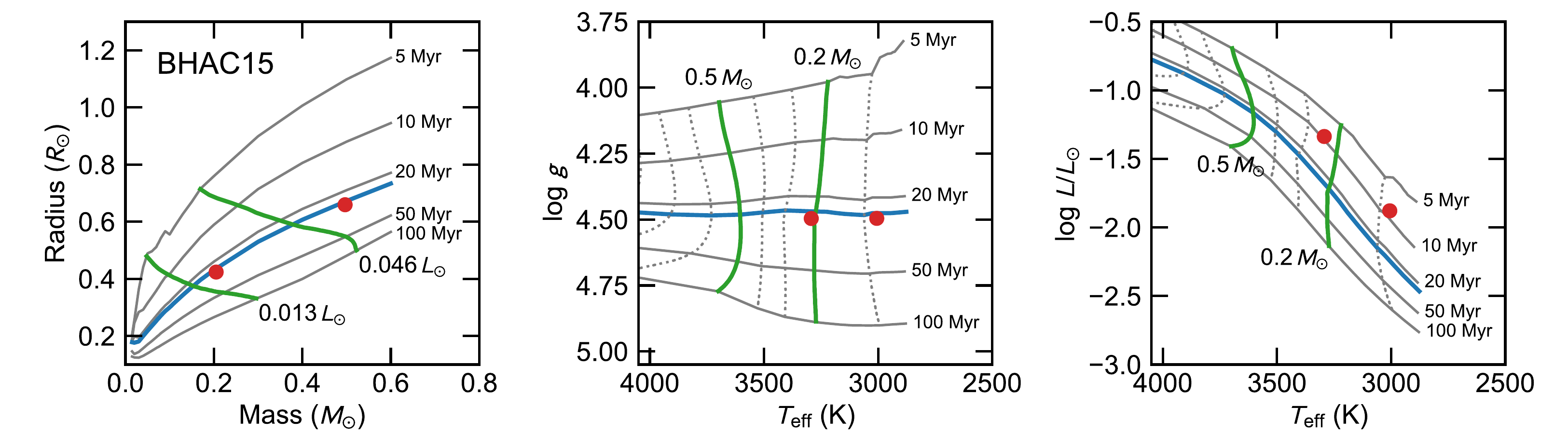}%
   \caption{Comparison of \eb\ against 5, 10, 20, 50 and 100 Myr  solar-metallicity \citet{Baraffe15} models. From left to right are the mass--radius, $T_{\rm eff}$--$\log g$ and $T_{\rm eff}$--luminosity (H--R diagram) planes. The thick blue line in each panel is the 24 Myr isochrone appropriate for members of the 32 Ori Group, while the thick green lines in the $T_{\rm eff}$--$\log g$ and $T_{\rm eff}$--luminosity plots are evolutionary tracks for the 0.2 and 0.5\,\Msun\ components of \eb. The green lines in the mass--radius plot are isolums at the \eblump\ and \eblums\,\Lsun\ we find for the primary and secondary components, respectively.}
   \label{fig:models_BHAC15}
\end{figure*}

\begin{figure*}
   \centering
   \includegraphics[width=\linewidth]{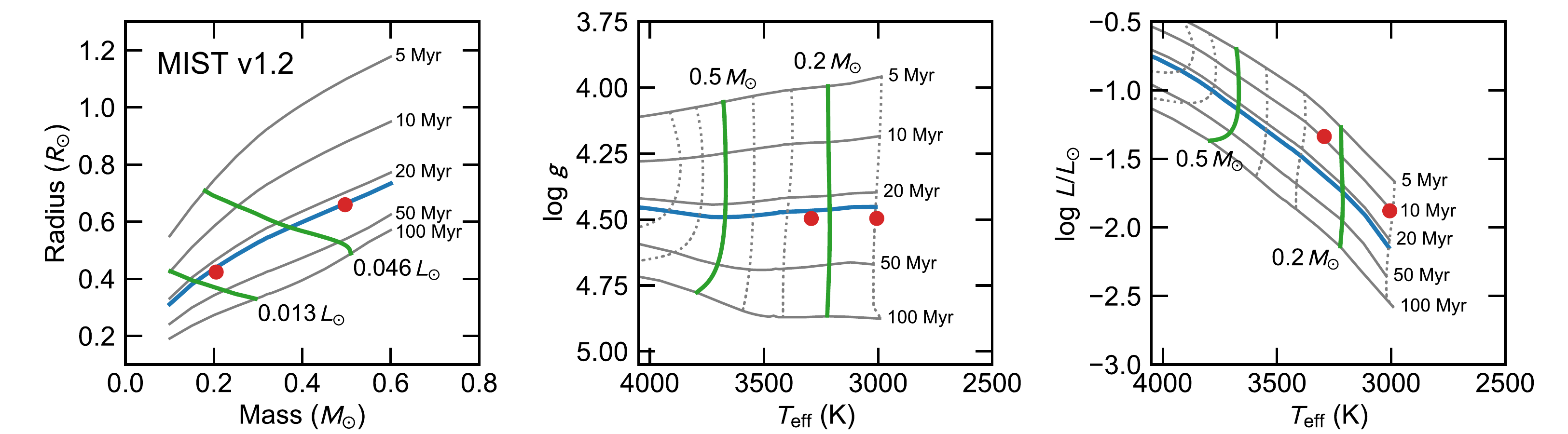}%
   \caption{As in \autoref{fig:models_BHAC15}, but for MIST v1.2 models \citep[$v/v_{\rm crit}=0.4$;][]{Dotter16,Choi16}.}
   \label{fig:models_MIST}
\end{figure*}

\begin{figure*}
   \centering
   \includegraphics[width=\linewidth]{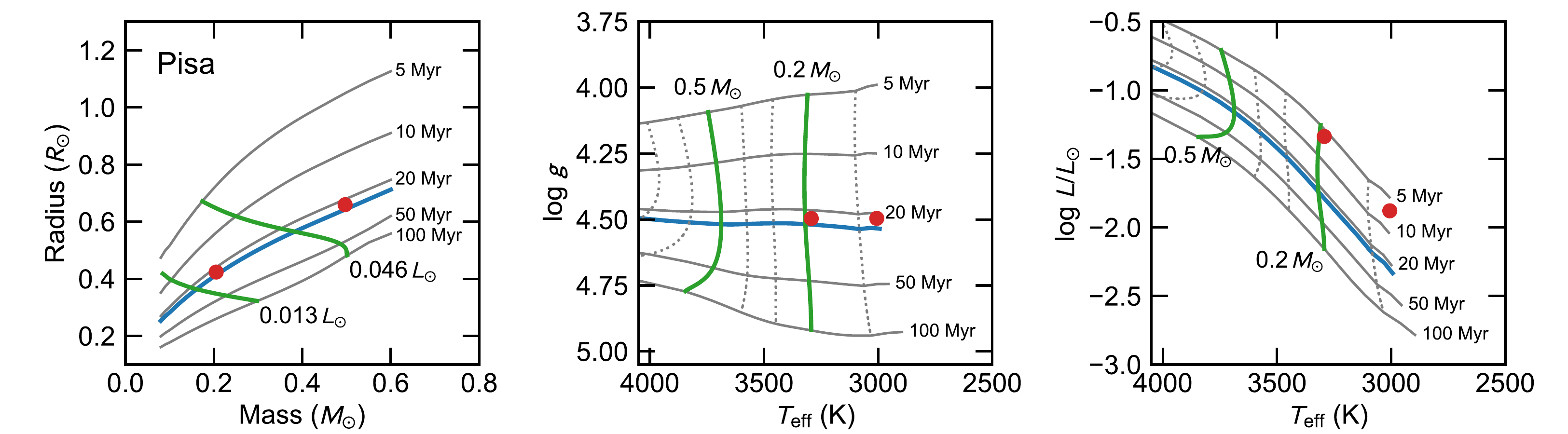}%
   \caption{As in \autoref{fig:models_BHAC15}, but for Pisa models \citep{Tognelli11}, extended down to 0.08\,\Msun.}
   \label{fig:models_Pisa}
\end{figure*}

\begin{figure*}
   \centering
   \includegraphics[width=\linewidth]{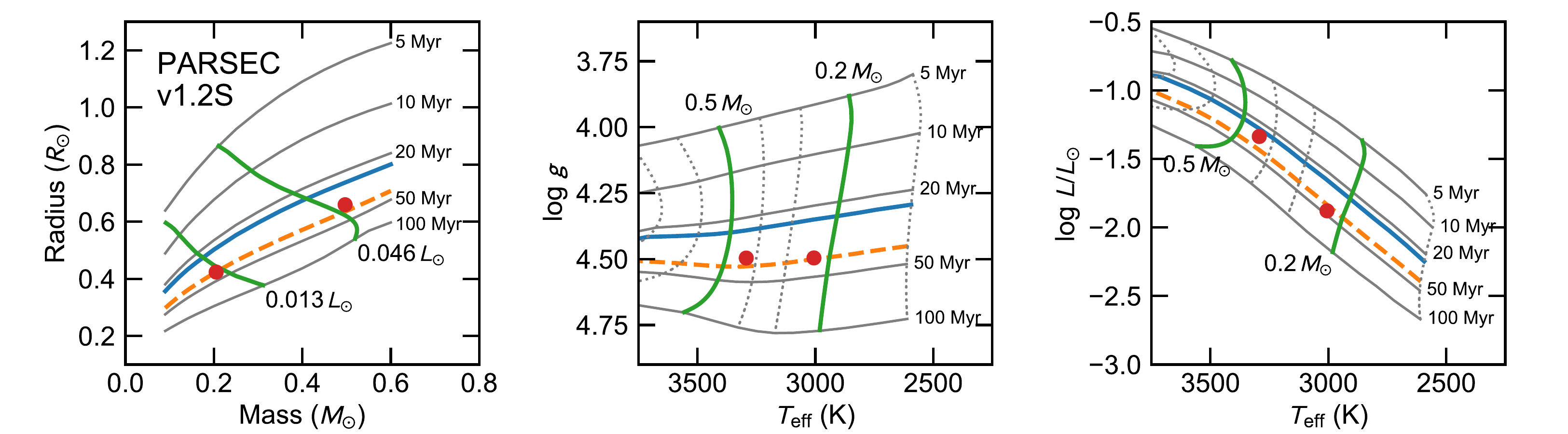}%
   \caption{As in \autoref{fig:models_BHAC15}, but for PARSEC v1.2S models \citep{Bressan12,Chen14}. The dashed yellow line in each panel is the 40 Myr isochrone.}
   \label{fig:models_PARSEC}
\end{figure*}

\begin{figure*}
   \centering
   \includegraphics[width=\linewidth]{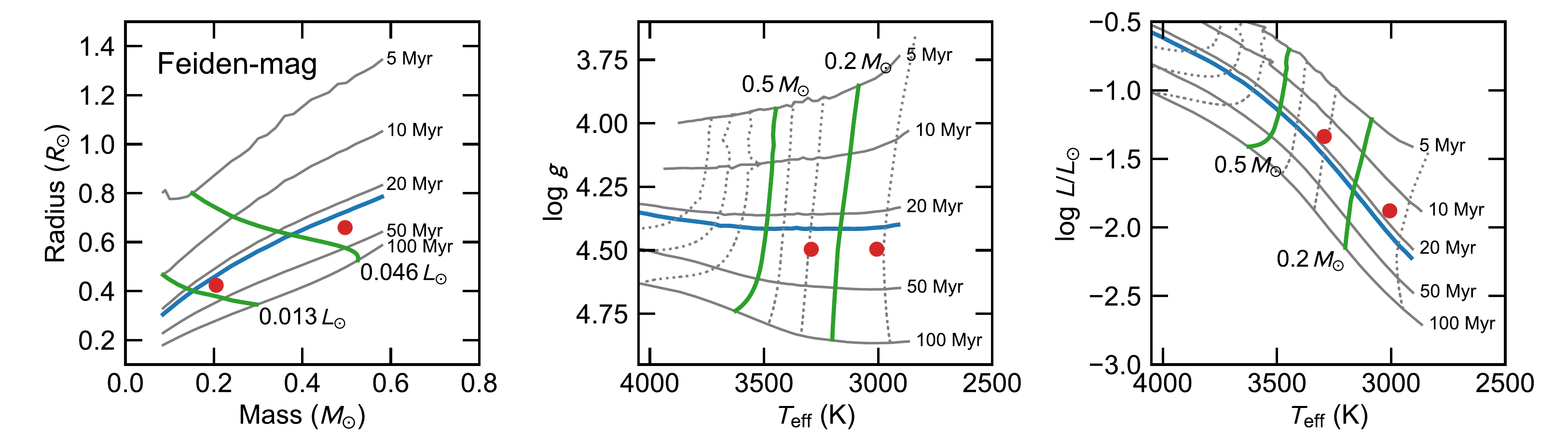}%
   \caption{As in \autoref{fig:models_BHAC15}, but for the solar-metallicity magnetic models of \citet{Feiden16}.}
   \label{fig:models_Feiden}
\end{figure*}

With its small mass ratio, well-determined parameters and young age, \eb\ permits precise comparisons to  evolutionary models across a mass and age range poorly constrained by observations. In Figs.\,\ref{fig:models_BHAC15}--\ref{fig:models_PARSEC} we plot the fundamental parameters of the system compared to several widely-used model grids: BHAC15 (\autoref{fig:models_BHAC15}), the MESA Isochrones and Stellar Tracks \citep[MIST \textsc{v}1.2, with $v/v_{\rm crit}=0.4$;][\autoref{fig:models_MIST}]{Choi16,Dotter16}, the Pisa models of \citet{Tognelli11} (extended down to 0.08\,\Msun; \autoref{fig:models_Pisa}) and version 1.2S of the PARSEC models \citep[][\autoref{fig:models_PARSEC}]{Bressan12} which include modified $T-\tau$ relations from BT-Settl model atmospheres as surface boundary conditions \citep{Chen14}. In all cases we adopt the solar-metallicity models.\footnote{Note that the grids adopt slightly different helium ($Y$) and heavy element ($Z$) fractions for their solar metallicity models, with BHAC15 using $(Y,Z)=(0.28,0.0153)$, MIST using $(0.2703,0.0143)$ and Pisa using $(0.253,0.013)$. For PARSEC we adopt the  $(Y,Z)=(0.273,0.014)$ tracks.}  The left panel of each figure shows that the components follow the expected gradient in the mass--radius diagram (MRD) and, with the exception of the PARSEC models, have radii close to those predicted for 24~Myr-old stars.  The interpolated MRD ages for each component are given in \autoref{table:ages}.   Uncertainties were calculated by a Monte Carlo simulation of $10^5$ normal samples considering the tabled uncertainties in mass and radius. The components of \eb\ are therefore coeval within their 1$\sigma$ uncertainties for the BHAC15 and Pisa models, but differ by 3$\sigma$ (only 2.4~Myr) for MIST.  The PARSEC v1.2S models imply significantly older ages (35--40\,Myr) which are less coeval. However, in contrast to the other models, the PARSEC isochrones at these ages would predict luminosities close to those we have calculated for \eb.

\begin{table*}
\hfil%
\begin{minipage}{\linewidth}
   \caption{Isochronal ages of \eb\ in the mass--radius (MR) and Hertzsprung--Russell (HR) diagrams, with interpolated  masses from the HRD.}
   \begin{tabular}{lccccccccccc} 
\hline
Model & \multicolumn{3}{c}{MRD age} & \multicolumn{3}{c}{HRD age} & \hfill & \multicolumn{4}{c}{HRD mass} \\
& Primary & Secondary & $\Delta t$ & Primary & Secondary & $\Delta t$ & \hfill & Primary & $\Delta M$ & Secondary & $\Delta M$ \\
 & (Myr) & (Myr)& (|$\sigma$|)& (Myr) & (Myr) & (|$\sigma$|) & \hfill & (\Msun) & (per cent) & (\Msun) & (per cent)\\
\hline
BHAC15 & $25.4 \pm 0.6$ & $26.2 \pm 0.6$ & 1.0 & 	   $9.9 \pm 1.4$ & $8.6 \pm 0.9$ & 0.8 & 	\hfill &   0.24 & $-$51 & 0.09 & $-$58 \\
MIST v1.2 & $24.5 \pm 0.5$ & $26.9 \pm 0.5$ & 3.2 & 	   $9.4 \pm 1.3$ & $11.0^{a}$ &  & \hfill &	   0.25 & $-$51 & 0.11 & $-$49 \\
Pisa & $22.0 \pm 0.5$ & $22.4 \pm 0.5$ & 0.5 & 	   $5.8 \pm 0.7$ & $\sim$6.5 &  & \hfill &	   0.19 & $-$62 & $<$0.08$^{b}$ & $-$61 \\
PARSEC v1.2S & $35.1 \pm 0.8$ & $40.3 \pm 0.7$ & 4.6 & 	   $29.2 \pm 3.0$ & $46.2 \pm 4.4$ & 3.2 & 	   \hfill & 0.44 & $-$11 & 0.23 & $+$12 \\
Feiden (standard) & $25.5 \pm 0.6$ & $26.3 \pm 0.5$ & 1.1 & 	   $8.7 \pm 1.2$ & $\sim$9.0 &  & \hfill &	   0.22 & $-$55 & $<$0.09$^{b}$ & $-$56 \\
Feiden (magnetic) & $31.8 \pm 0.7$ & $31.4 \pm 0.6$ & 0.4 & 	   $16.2 \pm 2.0$ & $15.9 \pm 1.6$ & 0.1 & 	   \hfill & 0.31 & $-$37 & 0.12 & $-$41 \\
\hline
   \end{tabular}\\
$^{a}$ Temperature error bar falls outside the model grid. This will bias the age uncertainties. \\
$^{b}$ Secondary falls just outside the model grid. The mass is therefore an upper limit. Ages were estimated visually.
\label{table:ages}
\end{minipage}
\hfil%
\end{table*}

Given the fast rotation rate of the primary and the strong \halpha\ and X-ray emission we observed, we can assume  both components are magnetically active. This activity is believed to produce stars with lower effective temperatures and inflated radii, either through magnetic fields reducing the efficiency of convection in the stellar interior, reduction of the effective radiating surface due to high spot coverage, or more likely a combination of both phenomena  \citep[e.g.][]{Chabrier07,MacDonald14,Feiden14,Somers15,Feiden16}.  \citet{Stassun12} presented empirical correlations for quantifying the temperature decrement and radius inflation observed in $<$1\,\Msun\ stars as a function of EW(\halpha) or the fractional X-ray luminosity. From both the (unresolved) $\log(L_{\rm X}/L_{\rm bol}) \approx -2.8$ we estimated in \autoref{sec:rotation} and the primary's $\textrm{EW(\halpha)}\approx\ -5$\,\AA, the relations predict a radius inflation of $\sim$14~per cent, with a corresponding temperature decrement of $\sim$6~per cent ($\sim$200~K for the 3300~K primary).  Assuming the age of \eb\ is not significantly older than $\sim$24~Myr, the excellent agreement between the measured radii and those predicted by the non-magnetic BHAC15, MIST and Pisa models for such an active system is challenging to explain. 

The effects of adding magnetic fields are illustrated in \autoref{fig:models_Feiden}, where we compare \eb\ to the models of \citet{Feiden16}.  These were computed assuming equipartition between the magnetic field pressure and gas pressure at a mean opacity of $\tau_{\textrm{5000\,\AA}}=1$ and adopt a surface magnetic field strength at each mass equal to the value at 10 Myr. Inhibition of convection by magnetic fields cools the stellar surface,  slowing a star's contraction and making it appear older at fixed mass than the non-magnetic models. This is indeed what we see in \autoref{fig:models_Feiden}, where the components of \eb\ remain coeval but have an inferred age of $\sim$32~Myr. Note that the surface gas pressure increases as the star contracts, so we expect the field strengths to be larger in models calibrated at 20--30~Myr. In contrast, the non-magnetic \citet{Feiden16} isochrones yield ages of $\sim$25~Myr, similar to the other models in \autoref{table:ages}.

Combining radius and mass to form the surface gravity, we see in the middle panels of Figs.\,\ref{fig:models_BHAC15}--\ref{fig:models_Feiden} that all the models are able to maintain coevality in the $T_{\rm eff}$--$\log g$ plane, with BHAC15 and MIST slightly overestimating the system age and the Pisa models underestimating it. Again, the PARSEC and magnetic models imply older ages. However, the evolutionary tracks for 0.5 and 0.2\,\Msun\ stars in these diagrams do not intersect the measured $\log g$ and $T_{\rm eff}$ (although the PARSEC tracks come close). The BHAC15, MIST and Pisa models all significantly underestimate the masses of the components. As extreme examples, consider the BHAC15 and Pisa models, which place the 0.5\,\Msun\ primary component on or below the 0.2\,\Msun\  track appropriate for the secondary.  

\begin{figure}
   \centering
   \includegraphics[width=\linewidth]{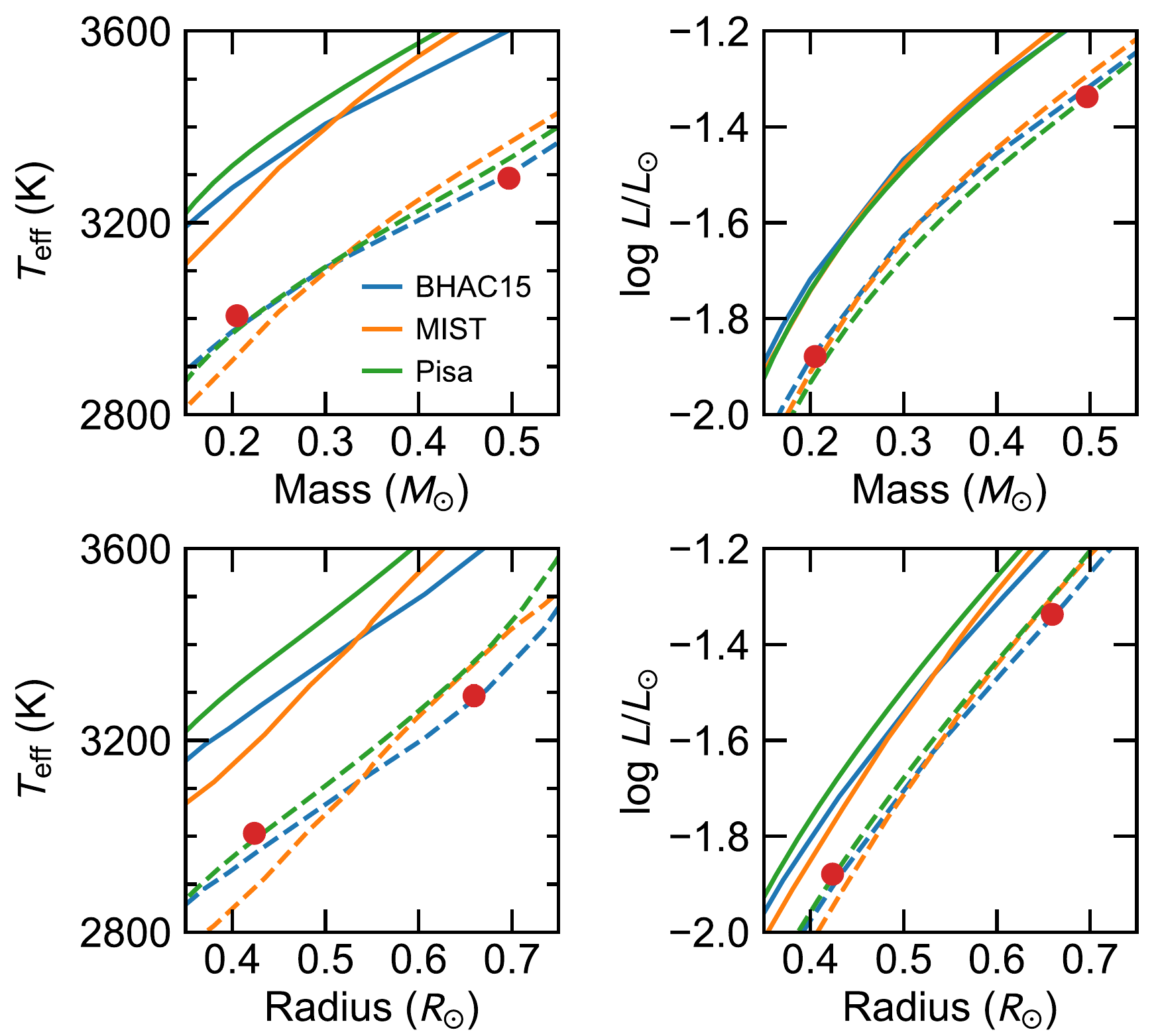}%
   \caption{Temperatures and luminosities of 24 Myr BHAC15, MIST and Pisa magnetic models as a function of mass and radius (solid lines). The red points are our measurements for \eb. Dashed lines show the effect of a simple shift in $T_{\rm eff}$ (or corresponding shift in $\log L$) of $-$300 K (BHAC15, MIST) or $-$350 K (Pisa) at fixed mass and radius.}
   \label{fig:teff_shift}
\end{figure}

Given the excellent agreement between the models and observations in the mass--radius plane, these discrepancies point to a problem with either the $T_{\rm eff}$ inferred from the SED fitting or the model temperature scales.  The SED fit incorporates strong constraints on the component masses and radii from the joint modelling, which should be more accurate than those derived from single star spectral type--$T_{\rm eff}$ relations \cite[e.g.][]{Pecaut13b,Herczeg14} applied to the primary star and transferred to the secondary through bandpass-dependent surface brightness ratios \citep[see][]{Gillen17}. Assuming the temperature scales of the BT-Settl and PHOENIX model atmospheres are correct, the BHAC15, MIST and Pisa tracks would need to be shifted approximately 300~K cooler to match the observations (\autoref{fig:teff_shift}). This is similar to the $\sim$200~K decrement predicted from the \citet{Stassun12} activity relations (which also used non-magnetic models). The effect is also visible in the mass--radius plane, where the models predict lower-than-observed radii at fixed mass and luminosity, a consequence of the overestimated temperatures.   Temperature shifts of similar size and direction have been reported across a variety of young, low-mass EBs and models  \citep[e.g.][]{Kraus15,David16a,David16,Gillen17,David19,Simon19},  suggesting that the models themselves are to blame. With cooler temperatures but uninflated radii, \eb\ is the inverse of the similarly-aged MML 53, where \citet{Gomez-Maqueo-Chew19} reported its components were larger but not cooler than predicted by non-magnetic evolutionary models.

The disagreement between theory and observation is even starker when moving to the more commonly-used H--R diagram (HRD) of $T_{\rm eff}$ versus $\log L$ (Figs.\,\ref{fig:models_BHAC15}--\ref{fig:models_Feiden}; right hand panels). Placing  the components of \eb\ on this diagram we infer (generally coeval) ages 2--4 times younger than from the mass--radius diagram (see \autoref{table:ages}). Again, shifting the models 300--350~K cooler (with a corresponding decrease in the Stefan--Boltzmann luminosity) does much to ameliorate the discrepancies (\autoref{fig:teff_shift}). However, this is unlikely to completely solve the problem, since the models cannot be shifted purely in $T_{\rm eff}$. Lower temperatures mean less energy radiated from the surface, slowing the contraction rate and changing the radius and  luminosity at a given age \citep{Kraus15}. The PARSEC models show minimal difference between their MRD and HRD ages but are not as coeval as the other grids.  However, they are the only models to show any coevality \emph{between} the observed planes. The components of \eb\ can be well fitted in all three panels by a 35--40~Myr isochrone (dashed lines in \autoref{fig:models_PARSEC}).

As with the $T_{\rm eff}$--$\log g$ diagram, the over-predicted temperatures mean the models predict much lower masses for \eb\ than we found from the orbital solution. The BHAC15, MIST, Pisa and non-magnetic Feiden models all underestimate the dynamical masses by 50--60~per cent (\autoref{table:ages}). Similar trends have been reported in the literature for other young systems with well-established masses \citep[e.g.][]{Hillenbrand04,Kraus15,Simon19}. The magnetic models do slightly better ($-$40~per cent), but only the PARSEC tracks come close to the true values; underestimating the primary mass by only $\sim$10~per cent and overestimating the secondary by a similar amount. 

It is evident from these comparisons that none of the models are able to simultaneously predict the mass, radius, temperature and luminosity of \eb\ at the assumed 24~Myr age of the 32 Ori Group. The BHAC15, MIST and Pisa models can reproduce the expected radii in the MRD, but require a shift in their temperature scales of $\sim$300~K to match the HRD positions of the components. The PARSEC models predict significantly older ages in both the MRD and HRD, but are better able to replicate the observed masses without a temperature shift. These models include ad hoc corrections to the BT-Settl $T-\tau$ surface boundary conditions at the lowest temperatures to better match the observed mass--radius relation of low-mass dwarfs \citep{Chen14}. While this has been suggested as an over-correction in the pre-main sequence regime \citep[i.e. if the adjustments are necessary due to missing opacities which are more important at higher gravities, e.g.][]{Kraus15}, perhaps by the age of the 32 Ori Group, \eb\ is sufficiently close to the main sequence for this to be no longer the case.

\section{Conclusion}
\label{sec:conclusion}

We have presented a full characterization of the young, low-mass eclipsing binary \eblong\ (=THOR 42), which we confirm is a member of the $\sim$24~Myr-old 32 Orionis Moving Group.  We have modelled the light and radial velocity curves to derive precise system parameters, yielding component masses and radii of \ebmp\ and \ebms\,\Msun, and \ebrp\ and \ebrs\,\Rsun, respectively (mass and radius uncertainties of 1~per cent and 0.5~per cent). With components spanning the fully-convective boundary for M dwarfs, THOR 42 provides a stringent test of evolutionary models near the start of the main sequence, which is currently not well constrained by observations.  

Surprisingly for such a tight (0.859\,d period), synchronized ($\vsini = 38$\,\kms) system, the radii we measure are no larger than those predicted by most non-magnetic models for a 20--25~Myr star (i.e. no anomalous inflation), in excellent agreement with the canonical age of the 32~Ori Group. However, none of the models can simultaneously predict the observed mass, radius, temperature and luminosity of the components at this age. Specifically, the H--R diagram position of THOR 42 would lead to 50--60~per cent smaller masses and 2--4 times younger ages being estimated from most model tracks and isochrones. The latest PARSEC models \citep{Chen14} preserve coevality across the mass--radius and H--R diagrams and come closest to replicating the dynamical masses of the system, but require a significantly older age of 35--40~Myr. A re-examination of the 32~Ori Group's isochronal and lithium depletion ages in light of \gaia\ parallaxes is necessary to confirm whether this older age can be supported. During this work we also found that two proposed 32~Ori Group members; THOR 33AB and 34, are in fact more likely members of the $\beta$ Pictoris Moving Group.

The discovery and characterization of more high-precision touchstone systems like THOR 42 at a range of masses and ages is necessary to calibrate the models across the entire pre-main sequence. Nearby young moving groups like 32 Orionis and the older subgroups of the Sco-Cen OB association will no doubt be especially fertile  grounds for such work.

\section*{Acknowledgements}

We thank Pierre Maxted, Edward Gillen, Andrew Casey and Melissa Ness for fruitful discussions regarding model  fitting, and the referee for a prompt and thorough review of the manuscript. We acknowledge the traditional owners of the land on which SSO is located, the Gamilaraay people, and pay our respects to their elders past and present. We also acknowledge the generosity of the indigenous Hawaiian people to allow observations from Haleakal\=a. 

SJM acknowledges the support of a Vice Chancellor's Postdoctoral Fellowship at the University of New South Wales Canberra and a visiting fellowship at the Australian National University. CPMB acknowledges support from the European Research Council (ERC) under the European Union's Horizon 2020 research and innovation programme (grant agreement no. 682115).

 This work includes data gathered with the 6.5 meter Magellan Telescopes located at Las Campanas Observatory, Chile.  Australian access to the Magellan Telescopes was supported through the National Collaborative Research Infrastructure Strategy of the Australian Federal Government.
 
 This work has made use of observations from the Las Cumbres Observatory Global Telescope (LCOGT) network, granted through guaranteed time of the Australian National University. 

Part of this research was carried out at the Jet Propulsion Laboratory, California Institute of Technology, under a contract with NASA.
 
This material is based upon work supported by the National Science Foundation Graduate Research Fellowship Program under Grant No. (DGE-1746045). Any opinions, findings, and conclusions or recommendations expressed in this material are those of the authors and do not necessarily reflect the views of the National Science Foundation.
 
 This work has made use of data from the European Space Agency (ESA) mission
{\it Gaia} (\url{https://www.cosmos.esa.int/gaia}), processed by the {\it Gaia}
Data Processing and Analysis Consortium (DPAC,
\url{https://www.cosmos.esa.int/web/gaia/dpac/consortium}). Funding for the DPAC
has been provided by national institutions, in particular the institutions
participating in the {\it Gaia} Multilateral Agreement. 

The national facility capability for SkyMapper has been funded through ARC LIEF grant LE130100104 from the Australian Research Council, awarded to the University of Sydney, the Australian National University, Swinburne University of Technology, the University of Queensland, the University of Western Australia, the University of Melbourne, Curtin University of Technology, Monash University and the Australian Astronomical Observatory. SkyMapper is owned and operated by The Australian National University's Research School of Astronomy and Astrophysics.  

This work includes data collected by the \tess\ mission, which are publicly available from the Mikulski Archive for Space Telescopes (MAST). Funding for the \tess\ mission is provided by NASA's Science Mission directorate.

\emph{Software:} \textsc{python} (v3.7.2), \textsc{ipython} \citep[v6.5.0;][]{Perez07}, \textsc{astropy} \citep[v3.2.1;][]{Astropy-collaboration13}, \ellc\ \citep[v1.8.4;][]{Maxted16}, \eleanor\ \citep[v0.2.4;][]{Feinstein19}, \textsc{emcee} \citep[v2.2.1;][]{Foreman-Mackey13}, \textsc{PyAstronomy} (v0.13.0, \url{https://github.com/sczesla/PyAstronomy}), \textsc{matplotlib} \citep[v3.1.1;][]{Hunter07}, \textsc{scipy} (v1.3.0; \url{http://scipy.org}), \textsc{numpy} \citep[v1.17.0;][]{Oliphant06}.



\vspace{-4mm}
\bibliographystyle{mnras}
\bibliography{mnras_template}


\vspace{-2mm}
\bsp	
\label{lastpage}
\end{document}